\documentclass[longauth]{aa}
\usepackage{booktabs}
\usepackage{ulem}
\usepackage{graphicx}
\usepackage{natbib}
\bibpunct{(}{)}{;}{a}{}{,}
\usepackage{txfonts}
\usepackage{xcolor}
\usepackage{hyperref}
\usepackage{multicol}
\usepackage{subfig}
\usepackage{appendix}
\usepackage{multirow}
\usepackage{footnote}
\begin{document}

\title{COALAS III: The ATCA CO(1-0) look at the growth and death of \\ H$\alpha$ emitters in the Spiderweb protocluster at z=2.16}
\titlerunning{Molecular gas properties of HAEs in the Spiderweb protocluster at z=2.16}
\author{J. M. P\'erez-Mart\'inez\inst{1,2}
          \and H. Dannerbauer\inst{1,2}
          \and B. H. C. Emonts\inst{3} 
          \and J.R. Allison\inst{4}  
          \and J. B. Champagne\inst{5}
          \and B. Indermuehle\inst{6}  
          \and R. P. Norris\inst{6,7} 
          \and P. Serra\inst{8} 
          \and N. Seymour\inst{9}
          \and A. P. Thomson\inst{10} 
          \and C. M. Casey\inst{11}
          \and Z. Chen\inst{12, 1, 2, 13}
          \and K. Daikuhara\inst{14}
          \and C. De Breuck\inst{15}
          \and C. D'Eugenio\inst{1,2}
          \and G. Drouart\inst{9}          
          \and N. Hatch\inst{16}  
          \and S. Jin\inst{17,18}
          \and T. Kodama\inst{14}
          \and Y. Koyama\inst{19,20}
          \and M. D. Lehnert\inst{21}
          \and P. Macgregor\inst{6}
          \and G. Miley\inst{22}
          \and A. Naufal\inst{19,23}
          \and H. R\"ottgering\inst{22}
          \and M. S\'anchez-Portal\inst{24}
          \and R. Shimakawa\inst{25,26}
          \and Y. Zhang\inst{27,28,1,2}
          \and B. Ziegler\inst{29}   
          }

\institute{Instituto de Astrofísica de Canarias (IAC), E-38205 La Laguna, Tenerife, Spain
\email{jm.perez@iac.es}
\and Universidad de La Laguna, Dpto. Astrofísica, E-38206 La Laguna, Tenerife, Spain
\and National Radio Astronomy Observatory, 520 Edgemont Road, Charlottesville, VA 22903, USA
\and First Light Fusion Ltd., Unit 9/10 Oxford Pioneer Park, Mead Road, Yarnton, Kidlington 0X51QU, UK
\and Steward Observatory, University of Arizona 933 N Cherry Ave, Tucson, AZ 85719, USA
\and Australia Telescope National Facility, CSIRO Space and Astronomy, PO Box 76, Epping, NSW 1710, Australia
\and Western Sydney University, Locked Bag 1797, Penrith,NSW 2751, Australia
\and INAF – Osservatorio Astronomico di Cagliari, Via della Scienza 5, I-09047 Selargius (CA), Italy 
\and International Centre for Radio Astronomy Research, Curtin University, 1 Turner Avenue, Bentley, WA 6102, Australia
\and Jodrell Bank Centre for Astrophysics, University of Manchester, Oxford Road, Manchester M13 9PL, UK
\and Department of Astronomy, The University of Texas at Austin, 2515 Speedway Blvd Stop C1400, Austin, TX 78712, USA
\and School of Astronomy and Space Science, Nanjing University, Nanjing
210093, China
\and Key Laboratory of Modern Astronomy and Astrophysics, Nanjing University, Nanjing 210093, China
\and Astronomical Institute, Tohoku University, 6-3, Aramaki, Aoba, Sendai, Miyagi, 980-8578, Japan
\and European Southern Observatory, Karl–Schwarzschild–Straße 2, D-85748 Garching bei M\"unchen, Germany
\and School of Physics and Astronomy, University of Nottingham, University Park, Nottingham NG7 2RD, UK
\and Cosmic Dawn Center (DAWN), University of Copenaghen, Jagtvej 128, T$\mathring{a}$rn I, 2200, Copenhagen, Denmark
\and DTU Space, Technical University of Denmark, Elektrovej 327, DK2800 Kgs. Lyngby, Denmark
\and Dept. of Astronomical Science, The Graduate University for Advanced Studies, 2-21-1 Osawa, Mitaka, Tokyo 181-8588, Japan
\and Subaru Telescope, National Astronomical Observatory of Japan, National Institutes of Natural Sciences, 650 North A’ohoku Place, Hilo, HI 96720, USA
\and Centre de Recherche Astrophysique de Lyon, ENS de Lyon, Université Lyon 1, CNRS, UMR5574, F-69230 Saint-Genis-Laval, France
\and Leiden Observatory, Leiden University, PO Box 9513, NL-2300 RA Leiden, the Netherlands
\and National Astronomical Observatory of Japan, 2-21-1 Osawa, Mitaka, Tokyo 181-8588, Japan
\and Instituto de Radioastronom\'ia Milim\'etrica (IRAM), Av. Divina Pastora 7, N\'ucleo Central, E-18012 Granada, Spain
\and Waseda Institute for Advanced Study (WIAS), Waseda University, 1-21-1 Nishi-Waseda, Shinjuku, Tokyo 169-0051, Japan
\and Center for Data Science, Waseda University, 1-6-1 Nishi-Waseda, Shinjuku, Tokyo 169-0051, Japan
\and Purple Mountain Observatory, Chinese Academy of Sciences, 10 Yuanhua Road, Nanjing, 210023, China
\and School of Astronomy and Space Science, University of Science and Technology of China, Hefei, Anhui 230026, China
\and Department of Astronomy, University of Vienna, T\"urkenschanzstrasse 17, A-1180 Vienna, Austria
}
%
%
%
%
\abstract
{We obtain CO(1-0) molecular gas measurements with the Australia Telescope Compact Array on a sample of 43 spectroscopically confirmed H$\alpha$ emitters in the Spiderweb protocluster at $\mathrm{z=2.16}$ and investigate the relation between their star formation activities and cold gas reservoirs as a function of environment. We achieve a CO(1-0) detection rate of $\mathrm{\sim23\pm12\%}$ with 10 dual CO(1-0) and H$\alpha$ detections within our sample at $\mathrm{10<\log M_{*}/M_\odot<11.5}$. In addition, we obtain upper limits for the remaining sources. In terms of total gas fractions ($\mathrm{F_{gas}}$), we find our sample is divided into two different regimes mediated by a steep transition at $\mathrm{\log M_{*}/M_\odot\approx10.5}$. Galaxies below that threshold have gas fractions that in some cases are close to unity, indicating that their gas reservoir has been replenished by inflows from the cosmic web. However, objects at $\mathrm{\log M_{*}/M_\odot>10.5}$ display significantly lower gas fractions than their lower stellar mass counterparts and are dominated by AGN (12 out of 20). Stacking results yield $\mathrm{F_{gas}\approx0.55}$ for massive emitters excluding AGN, and $\mathrm{F_{gas}\approx0.35}$ when examining only AGN candidates. Furthermore, depletion times of our sample show that most H$\alpha$ emitters at $\mathrm{z=2.16}$ will become passive by $\mathrm{1<z<1.6}$, concurrently with the surge and dominance of the red sequence in the most massive clusters. Our environmental analyses suggest that galaxies residing in the outskirts of the protocluster have larger molecular-to-stellar mass ratios and lower star formation efficiencies than galaxies residing in the core. However, star formation across the protocluster structure remains consistent with the main sequence, indicating that galaxy evolution is primarily driven by the depletion of the gas reservoir towards the inner regions. We discuss the relative importance of in-/outflow processes in regulating star formation during the early phases of cluster assembly and conclude that a combination of feedback and overconsumption may be responsible for the rapid cold gas depletion these objects endure.}
\maketitle
%
%
\section{Introduction}
\label{S:Intro}
The epoch known as the "cosmic noon" marks the peak of star formation and AGN activity in the Universe (\citealt{Madau14}). During this era, galaxy formation is largely driven by cold gas inflows from the surrounding structures (\citealt{Dekel09a}), and the star formation is regulated by such gas feeding as well as by gas outflows (feedback) due to supernova explosions and/or AGNs (\citealt{Genzel14}). On top of this, in high-density regions, there are more complexities introduced by the environmentally-driven physical processes. According to current cosmological models, the channeling of cold gas streams towards the center of galaxies, the effect of dynamical friction, and the onset of environmental effects contribute to compressing/altering the gas distribution (\citealt{Tacchella16}) of protoclusters members, possibly triggering central starburst (\citealt{Gomez-Guijarro19}) and supporting their inside out mass growth (\citealt{vanDokkum15}). Young forming high-z protoclusters are observed as strong overdensities of gas-rich, often dusty, star-forming and starbursting galaxies (\citealt{Dannerbauer14,Dannerbauer17}; \citealt{Casey16,Casey17};  \citealt{Calvi23}; \citealt{Toshikawa24}) as simulations also predict (e.g., \citealt{Chiang17}; \citealt{Remus23}). This galaxy population is believed to be the predecessors of the red ellipticals that start dominating the cores of the clusters by $z=1$ (e.g., \citealt{Ivison13}; \citealt{Smail14}). Thus, the rapid growth of galaxies within protoclusters is driven by the efficient transformation of the cold gas reservoir into stars until the sudden shutting down of star formation. However, the exact mechanisms terminating this phase of accelerated galaxy evolution remain unsettled, with both environmental effects and AGN feedback likely playing a role in the emergence of the first early-type galaxies in overdense environments (e.g., \citealt{Krishnan17}; \citealt{Polletta21}; \citealt{Mei23}).

The last two decades have seen a growing number of works trying to unveil the interplay between the protocluster environment and the accelerated evolution of the galaxy populations therein in terms of star-formation and AGN activity, metal enrichment, and molecular gas content (see \citealt{Overzier16} and \citealt{Alberts22} for a review). However, a unified picture depicting how the cluster assembly affects the early evolution of galaxies is still missing, with contradicting results between protoclusters at similar cosmic epochs probably due to the diversity of dynamical states that characterize these large-scale structures at the cosmic noon. In terms of star formation, several authors reported a higher activity within the most overdense regions at $\mathrm{z>2}$ thus suggesting the reversal of the so-called star formation density relation mediated by higher gas accretion and merging rates (e.g., \citealt{Alberts14}; \citealt{Shimakawa18a}; \citealt{Lemaux22}; \citealt{Monson21}; \citealt{Shi24a}: \citealt{PerezMartinez24}; \citealt{Laishram24}; \citealt{Taamoli24}). However, others report no such differences (e.g., \citealt{Toshikawa14}; \citealt{Cucciati14}; \citealt{Shi21}; \citealt{Sattari21}; \citealt{PerezMartinez23,PerezMartinez24}) in coeval protoclusters. A similar situation is found regarding the metal enrichment of galaxies with some authors claiming that the early development of an intracluster medium (ICM) in massive protoclusters would contribute to detaching infalling galaxies from the cosmic web, progressively hampering their gas accretion and suppressing their outflows, thus enriching their ISM (e.g., \citealt{Kulas13}; \citealt{Shimakawa15}; \citealt{PerezMartinez23}). On the other hand, various levels of metallicity deficiency (e.g., \citealt{Valentino15}; \citealt{Chartab21}; \citealt{Sattari21}; \citealt{PerezMartinez24}) or no significant environmental dependence at all (e.g., \citealt{Tran15}; \citealt{Kacprzak15}; \citealt{Namiki19}) have been reported in other overdense structures at similar redshifts. 

Furthermore, the slowdown of cold inflows in massive haloes is expected to take place over long periods (\citealt{Zolotov15}), contradicting observational evidence showing that massive galaxies at $z>1.5$ quench on much shorter timescales (\citealt{Barro16}). Therefore, the reduction and eventual shutting down of star formation in massive galaxies is unlikely to occur solely because accretion is fading away. Instead, AGN feedback could provide an effective and complementary route to quenching massive protocluster galaxies after their starburst phase by expelling out a significant fraction of their gas reservoir via vigorous galactic winds and by directly heating the interstellar medium. Interestingly, several works have suggested that the AGN fraction is enhanced in protoclusters at $\mathrm{2<z<4}$ (see \citealt{Lehmer09,Lehmer13}; \citealt{Krishnan17}; \citealt{Vito20,Vito24}; \citealt{Polletta21}; \citealt{Monson21}). Furthermore, recent deep X-ray observations hinted that the massive and partially virialized Spiderweb protocluster at $z=2.16$ may host up to 6 times higher AGN fraction than the coeval field ($\mathrm{\sim25\%}$ at $\mathrm{M_*>10^{10.5}M_\odot}$, \citealt{Tozzi22a}). By contrast, less evolved protoclusters in terms of their stellar-mass build-up (e.g. USS1558 at $z=2.53$, \citealt{Shimakawa18a}; \citealt{PerezMartinez24}; \citealt{Daikuhara24}) do not show such enhancement of X-ray sources at similar depth (\citealt{Macuga19}). This suggests that the transformative processes triggering AGN activity (e.g., angular momentum loss of the gas and/or merging) have yet to occur in such young protoclusters thus reflecting the diversity among the evolutionary stage of these large-scale structures. Furthermore, $\mathrm{z\lesssim1}$ galaxies residing in virialized galaxy clusters display lower AGN activity than their coeval field counterparts (\citealt{Haines12}; \citealt{Ehlert14}; \citealt{Koulouridis19}; \citealt{Koulouridis24}), thus suggesting that the processes triggering the rapid supermassive black hole growth in cosmic noon protoclusters are short-lived and lose efficiency as these structures grow in mass and achieve virialization. 

This puzzling situation has proven that the environmental effects present in protoclusters, if any, are mildly reflected on the properties of the ionized gas phase, and are easily washed out by the intrinsic scatter of the studied scaling relations due to relatively poor number statistics (e.g., \citealt{Torrey19}; \citealt{Huang23}). In order to get deeper insights we should move to investigate the physical quantities that give birth to the scaling relations responsible for the mass growth and metal enrichment of galaxies: the cold molecular gas reservoir. The molecular gas properties of protocluster galaxies at $z>2$ are thus key to understanding the gas feeding and consumption processes fueling star formation and their changing efficiency as a function of both redshift, cluster assembly stage, and nuclear activity. Over the last years, there have been a growing number of works studying the molecular gas phase of star-forming protocluster galaxies using different CO transitions and dust continuum, although often over relatively small areas, sample sizes and focused on relatively bright sources (\citealt{Tadaki14}; \citealt{Dannerbauer14, Dannerbauer17}; \citealt{Emonts16,Emonts18}; \citealt{Rudnick17}; \citealt{Lee17}; \citealt{Coogan18}; \citealt{Wang18}; \citealt{Tadaki19}; \citealt{Zavala19}; \citealt{Gomez-Guijarro19}; \citealt{Champagne21}; \citealt{Jin21}; \citealt{Aoyama22}; \citealt{Zhang22}; \citealt{Polletta22}; \citealt{Ikeda22}; \citealt{Pensabene24}). Thus, it is important to expand these works to investigate the relatively mainstream star-forming population at this redshift range (i.e., main sequence galaxies) and cover a wider area within protoclusters to disentangle possible early environmental effects from other internal self-regulating ISM processes. However, very few fields have met these requirements up to now.

Over the last decade, a growing number of studies have identified star-forming (or star-bursting) protoclusters in the early universe with various techniques. The Spiderweb protocluster at $\mathrm{z=2.16}$ is a spectacular example of such high-z structures linked to a central radio galaxy MRC\,1138--262 (so-called "Spiderweb"; \citealt{Miley06}). This structure was originally discovered as an overdensity of Ly$\alpha$ emitters (LAEs, \citealt{Pentericci00}; \citealt{Kurk00}), and has gathered multiwavelength follow-up observations exceeding $>1200$ hours over the last 25 years including X-ray and $UgRIzJHK_s$ photometry, as well as IR-to-millimetre mapping with Spitzer, APEX-LABOCA, Herschel, AzTEC, ALMA, and ATCA, thus confirming the overdensity through a large variety of galaxy populations (e.g., \citealt{Kurk04}; \citealt{Croft05}; \citealt{Kodama07}; \citealt{Zirm08}; \citealt{Tanaka10}; \citealt{Doherty10};  \citealt{Kuiper11}; \citealt{Hatch11}; \citealt{Rigby14}; \citealt{Shimakawa14,Shimakawa15,Shimakawa18a,Shimakawa24}; \citealt{Dannerbauer14,Dannerbauer17};  \citealt{Zeballos18}; \citealt{Tadaki19}; \citealt{Emonts16,Emonts18}; \citealt{Jin21}; \citealt{DeBreuck22}; \citealt{Carilli22}; \citealt{Tozzi22a,Tozzi22b}; \citealt{Anderson22}; \citealt{PerezMartinez23}; \citealt{ChenZ24}; \citealt{Naufal23}; \citealt{Lepore24}; \citealt{Daikuhara24}). In particular, a deep and wide-field ($100$\,cMpc$^2$) narrow-band search within this field, as part of the MAHALO-Subaru project (\citealt{Kodama13}) yielded a huge ($\sim$10 Mpc scale) overdensity traced by H$\alpha$ emitters (HAEs, \citealt{Koyama13}) whose membership has been confirmed by recent spectroscopic follow-ups (\citealt{Shimakawa15}, \citealt{PerezMartinez23}). In addition, recent results from a narrow-band JWST/NIRCam cycle 1 campaign have unveiled a new population of $\mathrm{Pa\beta}$ emitters that overlaps with the spatial distribution of HAEs and whose dust attenuation has been explored (\citealt{Shimakawa24b}; \citealt{Perez-Martinez24b}). Furthermore, as part of the COALAS project, \cite{Jin21} revealed an overdensity of CO(1-0) emitters at $z=2.1-2.2$ hosting a large number of extended molecular gas reservoirs as shown by \cite{ChenZ24}. In addition, a large overdensity of bright dust-continuum emitting sources has been confirmed by \citealt{Zhang24} through ALMA observations. As a result, the dynamical mass of this protocluster exceeds $M_{\rm cl}\sim2\times10^{14}M_{\odot}$ (\citealt{Shimakawa14}). Recently, \cite{Tozzi22a} revealed the presence of hot, diffuse baryons on a scale of $\mathrm{\sim100\,kpc}$ from the Spiderweb galaxy using X-ray observations and \cite{DiMascolo23} reported signs of a nascent ICM via Sunyaev-Zeldovich effect implying this structure will soon become a bonafide galaxy cluster. Moreover, a handful of massive passive objects already populate the core of the protocluster (\citealt{Naufal24}), adding evidence to the mature stage of this large-scale structure. Thus, the Spiderweb protocluster stands out as the best-known example of a confirmed galaxy cluster in formation at the cosmic noon and as an ideal test site to search for the first environmental effects from a multiwavelength perspective.

In this work, we search for environmental effects acting over the star formation activities and molecular gas reservoirs of H$\alpha$ emitters belonging to the Spiderweb protocluster at $\mathrm{z=2.16}$. We structured this manuscript in the following way: $\text{Sect.}$\,\ref{S:Data} describes the observations and datasets used available within this field and part of our work. $\text{Sect.}$\,\ref{S:Methods} outlines the methods used to analyze the physical properties of our targets and the environmental parameters measured. $\text{Sect.}$\,\ref{S:Results} and \ref{S:Discussion} respectively present our main results and their physical interpretation in the context of galaxy evolution in overdense environments at the cosmic noon. Finally, $\text{Sect.}$\,\ref{S:Conclusions} outlines the major conclusions of this study. Throughout this article, we adopt a flat cosmology with $\Omega_{\Lambda}$=0.7, $\Omega_{m}$=0.3, and $H_{0}$=70 km\,s$^{-1}$Mpc$^{-1}$ and we assume a \citet{Chabrier03} initial mass function (IMF). Finally, all magnitudes quoted in this paper are in the AB system (\citealt{Oke83}).

\section{Observations and datasets}
\label{S:Data}

This section provides an overview of the datasets employed in our analyses, including their origins and key attributes. Our work relies on two main collections of data: the COALAS project \footnote{http://research.iac.es/proyecto/COALAS/pages/en/home.php} whose focus is to identify CO(1-0) emitters across the Spiderweb protocluster structure and investigate their properties and dependencies with the environment; and the MAHALO project (\citealt{Kodama13}), which aims at selecting narrow-band H$\alpha$ emitters in high-z overdensities and characterize them through subsequent spectroscopic follow-ups. Additionally, we leverage the extensive repository of archival multiwavelength broad-band photometric data available for the Spiderweb protocluster field. 

\subsection{COALAS}
\label{SS:COALAS} 

The COALAS project is a large program (ID: C3181, PI: H. Dannerbauer) conducted with the Australian Telescope Compact Array (ATCA). Its primary objective is to explore into the influence of cosmic environments on the molecular gas reservoirs of galaxies in different environments during the cosmic noon, specifically through the ground-state transition of CO (i.e., $J=1$). Furthermore, the CO(1-0) transition is the most robust tracer of the cold molecular gas and has been extensively used in the past for this purpose even at $z>1.5$ (e.g., \citealt{Dannerbauer09}; \citealt{Carilli10}; \citealt{Ivison11}; \citealt{Emonts13}). The observations used in this work (i.e., approximately half of the COALAS project) took place from April 2017 to March 2020 targetting the central regions of the Spiderweb protocluster with 13 overlapping pointings, thus forming a mosaic of approximately 25.8 $\mathrm{arcmin^2}$ (Fig.\ref{F:MAP}). Throughout this program, approximately 475 hours of integration time were accumulated in this field with a central observed frequency of $\mathrm{36.5\,GHz}$ corresponding to the CO(1-0) emission line at $z=2.162$, and bandwidth of $\mathrm{2\,GHz}$ spanning $\mathrm{\pm7000\,km/s}$ in velocity space across the line of sight of the protocluster in channels of $\mathrm{\pm40\,km/s}$ each and with a pixel size of $1.5\arcsec$. Due to the variety of array configurations and integration times between pointings, the measured rms and beam size significantly vary across the mosaic. The typical rms level per beam for individual pointings ranges at $\mathrm{0.13-0.29\,mJy}$ per channel, while the maximum and minimum beam sizes found in the mosaic are $\mathrm{4.8\arcsec\times3.5\arcsec}$ and $\mathrm{13.8\arcsec\times13.1\arcsec}$, which corresponds to roughly 30\,kpc and 100\,kpc across respectively. Because of this low spatial resolution, the peak flux of a source is representative of the total flux of such an object. We refer to Table 1 in \cite{Jin21} for further details on the array configurations for each individual pointing and data reduction details. In total, \cite{Jin21} identified 46 CO(1-0) emitters at $\mathrm{S/N>4}$ forming an overdensity that spans a redshift range of $z=2.12-2.21$, thus suggesting the presence of a large-scale filamentary structure surrounding the Spiderweb protocluster or a galaxy superprotocluster similar to the recently discovered Hyperion at $z=2.4$ (\citealt{Cucciati18}) or Elentári at $z=3.3$ (\citealt{Forrest23}). The probability of false CO(1-0) line detection for these sources as defined by \cite{Jin21} is typically $\mathrm{<10\%}$. In this work, we will use the COALAS mosaic to explore the evolution of molecular gas reservoirs in star-forming galaxies residing in the Spiderweb protocluster.

\subsection{MAHALO-Subaru}
\label{SS:MAHALO} 

\begin{figure*}    
\centering
\includegraphics[width=14cm]{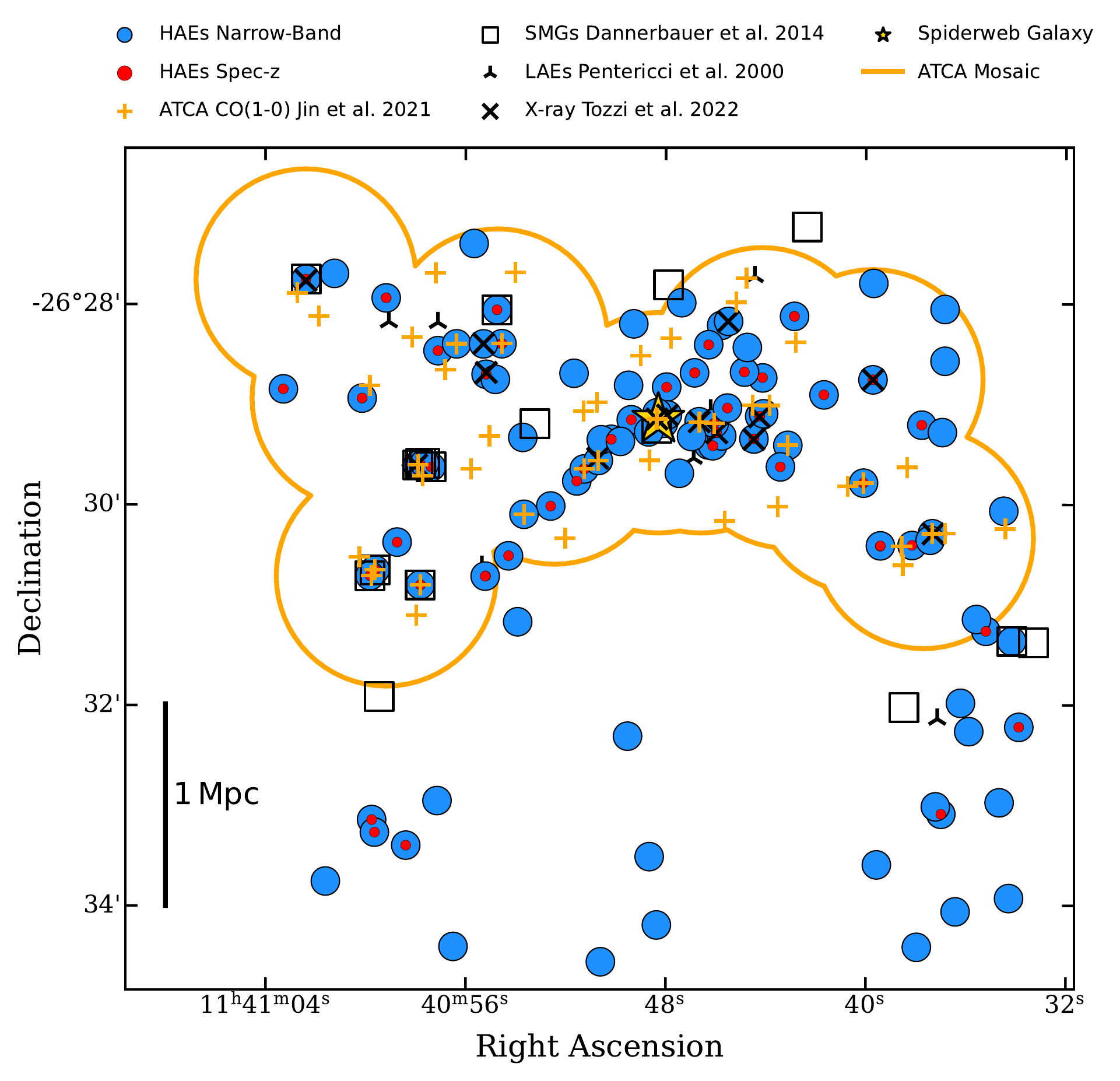}\par 
\caption{Spiderweb protocluster field at $\mathrm{z\approx2.16}$. Blue circles display the full sample of candidate HAEs from \citet{Koyama13} and \citet{Shimakawa18b}. Red circles show those HAEs with measured spectroscopic redshift (\citealt{Shimakawa18b} and \citealt{PerezMartinez23}). Orange crosses and contours respectively show the CO(1-0) emitters reported by \citet{Jin21} and the limits of the COALAS ATCA footprint. Empty squares depict the subsample of SMGs reported by \citet{Dannerbauer14} in this field. Black stars show the LAEs from \cite{Pentericci00}, while X-ray sources from \citet{Tozzi22a} are shown by black crosses. The Spiderweb galaxy is marked with a yellow star. We display a bar with 1 Mpc physical scale at $\mathrm{z=2.16}$ in the lower-left corner of the diagram for reference.}
\label{F:MAP}
\end{figure*}

The Spiderweb protocluster field also has extensive multiwavelength spectrophotometric coverage from the MAHALO-Subaru project, which aims at investigating the evolution of the star-forming activities of galaxies residing in overdensities up to $z=2.5$ (\citealt{Kodama13}). In particular, the available datasets include Subaru/Suprime-Cam B and z$'$ bands, Subaru/MOIRCS J and $\mathrm{K_s}$-band and Subaru/MOIRCS narrow band observations with the filter NB2071 to identify HAEs (\citealt{Koyama13}; \citealt{Shimakawa18b}) across the Spiderweb protocluster structure. The MAHALO-Subaru project also conducted a series of spectroscopic follow-ups over the existing population of narrow-band selected HAEs yielding $>50$ spectroscopic redshift through past programs with SUBARU/MOIRCS (\citealt{Shimakawa14,Shimakawa15}) and VLT/KMOS (\citealt{PerezMartinez23}). In this work, we will use these spectroscopically confirmed protocluster HAEs as our parent sample. In addition, we will crossmatch the position of these sources with the ATCA mosaic from the COALAS project (\citealt{Jin21}) to obtain CO(1-0) molecular gas masses for some sources and upper limits for the rest. In total, the Spiderweb protocluster field has 54 spectroscopically confirmed narrow-band selected HAEs (\citealt{PerezMartinez23}; \citealt{Shimakawa24}) and the ATCA mosaic from the COALAS project overlaps with the positions of 49 of them as shown in Fig.\,\ref{F:MAP}. 

\begin{figure}    
\centering
\includegraphics[width=\linewidth]{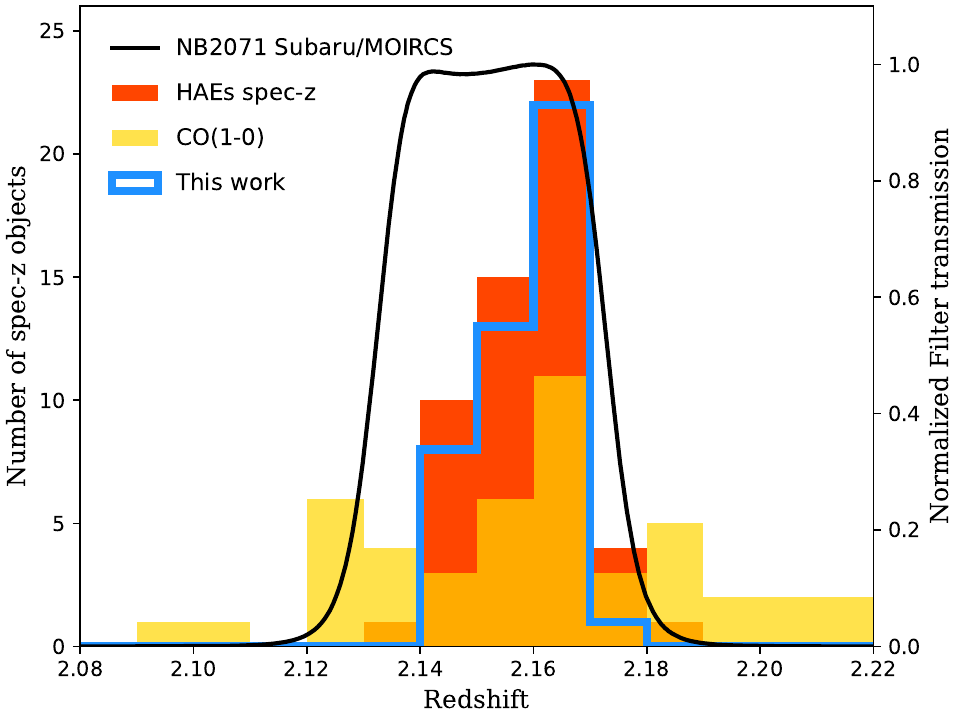}\par 
\caption{Redshift distribution of sources. The complete sample of spectroscopic HAEs (\citealt{Shimakawa18b,Shimakawa24}; \citealt{PerezMartinez23}) is shown in red. The sample of ATCA CO(1-0) emitters (\citealt{Jin21}) is shown in yellow. The sample of HAEs within the ATCA footprint (see Fig.\,\ref{F:MAP}) that is used in this work is depicted by the blue contour histogram. The black solid line shows the shape and redshift coverage of the Subaru/MOIRCS narrow-band filter used to select the HAEs.}
\label{F:Hist}
\end{figure}

Given its distinct nature, we exclude the Spiderweb Galaxy complex from our analyses. This is often referred to as the $\sim$100 kpc region within the giant Ly-alpha nebula that surrounds the radio galaxy MRC\,1138-262 and its satellite galaxies ("flies", \citealt{Kuiper11}). Through hierarchical merging, this region is likely destined to evolve into a single Brightest Cluster Galaxy (BCG) at low redshift (\citealt{Pentericci97}; \citealt{Hatch08,Hatch09}). In addition, we remove another five HAEs. Two of them (IDs 121 and 1185, \citealt{Koyama13}) lie in the noisy region close to the edge of the ATCA mosaic. Consequently, they are unsuitable for accurate CO(1-0) measurements (see \citealt{Jin21}). Another two HAEs (IDs 469 and 1106) were spectroscopically confirmed only by their CO(1-0) emission. However, these redshifts would position their $\mathrm{H\alpha}$ emission at the edge of NB2071, where the transmission is less than 10\%, rendering any attempts to measure $\mathrm{H\alpha}$ with the narrow-band filter unreliable. One additional source has a redshift mismatch of approximately $\mathrm{\Delta z \approx 0.04}$ between its $\mathrm{H\alpha}$ and CO(1-0) emissions, which complicates its redshift determination and leads to its exclusion from the analysis. Thus, the final sample of HAEs within the ATCA mosaic analyzed in this work consists of 43 sources. Figure \ref{F:Hist} presents the redshift distributions for the spectroscopic HAEs (\citealt{PerezMartinez23}; \citealt{Shimakawa24}), the CO(1-0) sources (\citealt{Jin21}), and the specific subsample used in this study.

\subsection{Multiwavelength broad-band photometry}
\label{SS:Photometry} 

In addition to datasets coming from the previous standalone projects, this field also counts with further NIR broadband deep imaging taken by VLT/HAWK-I in Y, H, $\mathrm{K_s}$ (PI: J. Kurk, program IDs 088.A-0754, 091.A-0106, 094.A-0104, see \citealt{Dannerbauer17} and \citealt{Shimakawa18b}). Furthermore, part of the field overlaps with broadband imaging at 3.6 and 4.5 $\mathrm{\mu m}$ made by Spitzer/IRAC (PI: D. Stern, campaign IDs 736 and 793, see \citealt{Seymour07}) whose Post-BCD (PBCD) products can be retrieved from the Spitzer Heritage Archive (SHA). In addition, the Hubble Legacy Archive provides a small mosaic of reduced Hubble Space Telescope (HST) ACS/WFC data for filters F475W and F814W (PI: H. Ford, proposal ID 10327, see \citealt{Miley06}). Finally, recent deep X-ray images with Chandra (PI: P. Tozzi, proposal ID: 20700463, see \citealt{Tozzi22a}) surveyed both the COALAS and MAHALO-Subaru field in the Spiderweb protocluster thus providing a list of X-ray emitting sources whose positions have been crossmatched with the known existing HAEs to pinpoint the location AGN candidates within the protocluster structure. We refer to the main publications accompanying these datasets for further details on their instrument configurations and depths.

\section{Methods}
\label{S:Methods}

\subsection{Molecular gas masses from CO(1-0)}
\label{SS:CO}

In this section, we use the COALAS datacube mosaic in the Spiderweb protocluster to estimate the molecular gas masses for the spectroscopically confirmed protocluster sample of 43 HAEs. We gather the CO(1-0) fluxes at $\mathrm{S/N\gtrsim4}$ reported by \cite{Jin21} for their sample of 14 dual $\mathrm{H\alpha+CO(1-0)}$ emitters. We remove the Spiderweb radio galaxy due to its distinct nature and three other sources from this list following Sect\,\ref{SS:MAHALO}. This results in a final sample of 10 dual $\mathrm{H\alpha+CO(1-0)}$ emitters from \cite{Jin21}. We need to recover new and reliable CO(1-0) flux upper limits for the remaining 33 spectroscopic HAEs across the datacube footprint. \cite{Jin21} based his CO(1-0) S/N measurements on the rms noise at various locations in the datacube. However, because the rms noise varies substantially between the pointings in the mosaic, and because we want to rule out that low-level instrumental effects across the ATCA band may affect our data (see \citealt{Emonts11}), we will adopt a two-step process to derive our upper limits based on the known sky coordinates and redshifts of these 43 HAEs (\citealt{Shimakawa18b}; \citealt{PerezMartinez23}).

First, we take the sky coordinates of the spectroscopic HAEs (\citealt{Koyama13}; \citealt{Shimakawa18a}; \citealt{PerezMartinez23}). The ATCA configuration during the COALAS project yielded a relatively wide but also variable beam size across the surveyed field. However, the typical distance between the H$\alpha$ and CO(1-0) peak emission of dual emitters is less than twice the spaxel size (i.e., $<3\arcsec$). Thus, we take grid of $\mathrm{3\times3}$ spaxels around each H$\alpha$ emitter's position as the tentative locations to measure its flux upper limits (i.e., peak flux). However, to decide which spaxel contains the highest peak flux we need to define a spectral window to integrate the CO(1-0) flux upper limit, and then repeat the process across the grid in an iterative way. 

\begin{figure}    
\centering
\includegraphics[width=\linewidth]{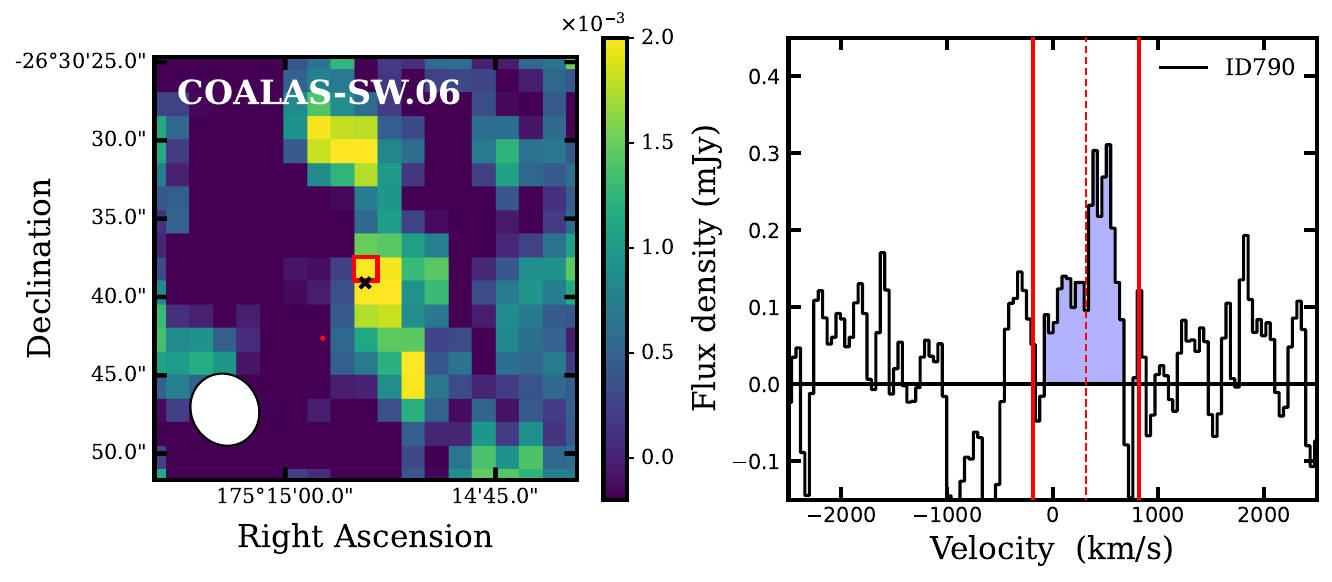}\par 
\includegraphics[width=\linewidth]{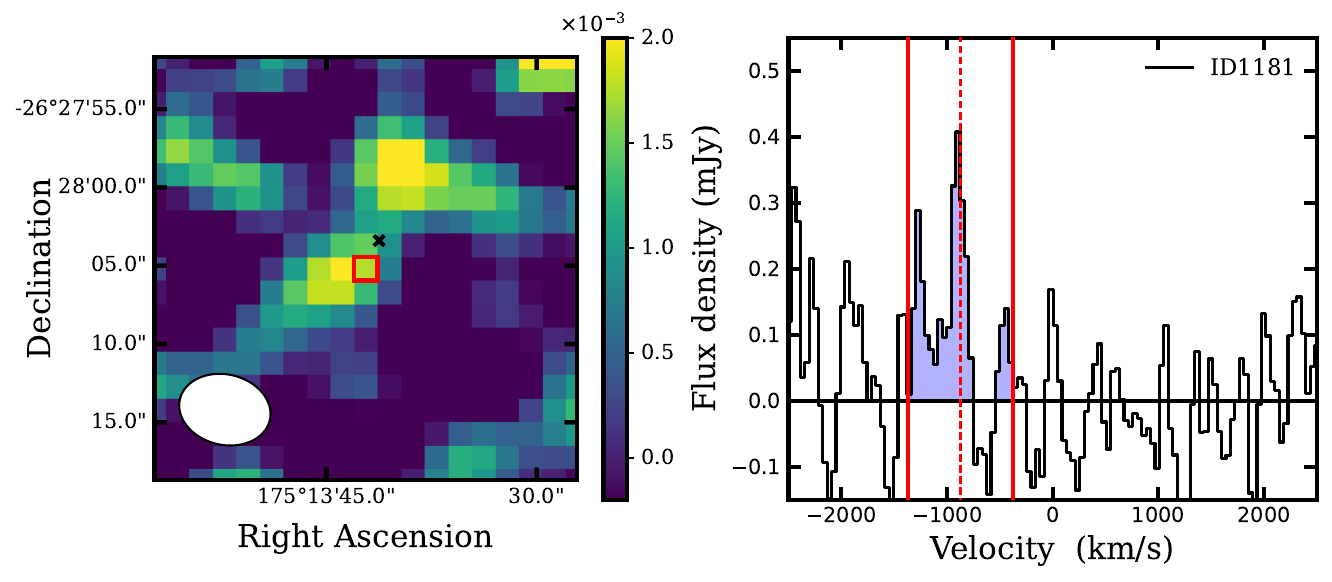}\par 
\caption{Two examples of our upper limit determination method. The upper panels display the HAE ID 790 for which \citet{Jin21} reported CO(1-0) at $\mathrm{S/N>4}$, while the lower panel displays an undetected CO(1-0) source (HAE ID 1181) for which we compute upper limits. The first column displays the moment zero of the ATCA datacube after collapsing it at the redshift of our source $\mathrm{\pm500\,km/s}$. The black cross shows the exact position of the HAE under scrutiny. The red empty square shows the position of the spaxel containing the peak flux after inspecting a $\mathrm{3\times3}$ grid around the HAEs spaxel. The beam size is shown in the lower left corner. The second column depicts the spectra extracted from the peak flux spaxel. The vertical red dashed line shows the systemic velocity (i.e., redshift) of the galaxy from $\mathrm{H\alpha}$ while the solid red lines display the limits ($\mathrm{\pm500\,km/s}$) for flux integration (blue area). Velocities are relative to $\mathrm{z=2.1612}$ as in \citet{Jin21}}.
\label{F:Examples}
\end{figure}

Thus, we use the spectroscopic redshifts measured for each HAE as a reference point in the spectral dimension to search for CO(1-0) emission. Then, we define a spectral window of $\mathrm{1000\,km/s}$ width around the reference redshift of each source. This number has been chosen by taking into account that the median full width at zero intensity (FWZI) of the COALAS sample is approximately $\mathrm{600\pm400\,km/s}$ (\citealt{Jin21}) and thus, our spectral window encompass the flux contained up to 1$\sigma$ beyond the median FWZI value. Clustering effects (i.e., signal overlapping) are only relevant in the vicinity of the Spiderweb complex given its bright and extended CO(1-0) emission. However, this part of the mosaic also has the best spatial resolution. Only two objects ID 1440 and 1501 would be affected by this situation and are considered as upper limits (see Table\,\ref{T:BigTable}), albeit they are separated from the systemic velocity of the Spiderweb galaxy by $\mathrm{\gtrsim2\times FWHM_{SW,CO(1-0)}}$ (\citealt{Jin21}), implying that strong contamination is unlikely.

Finally, our approach integrates only the positive flux channels within the spectral window, thus guaranteeing that we are not underestimating the measured flux for our upper limits. Last, we repeat this process for all the spaxels within the $\mathrm{3\times3}$ spatial grid centered on the peak position of the H$\alpha$ emission and select the final upper limit for each source as the highest recovered flux between these 9 spaxels. In Fig.\,\ref{F:Examples}, we show two examples of this method applied to a source with CO(1-0) detection at $\mathrm{S/N>4}$ in \cite{Jin21} and to a source below this threshold (i.e., not reported) in the same work. We achieve a median $\mathrm{S/N\approx2.3}$ for our upper limits following the prescription given by Eq.\,1 in \cite{Jin21} and assuming $\mathrm{FWZI=1000\,km/s}$, which represents the width of our spectral window. Upper limits at $\mathrm{S/N<1}$ are discarded and we take the rms value itself as the final flux estimate. We summarize our upper flux limits measurements among other properties of our sample in the appendix (Table\,\ref{T:BigTable}). Following our method, there are 4 cases (IDs 229, 999, 1139, 1385) out of the 33 spectroscopic HAEs in this field for which we recover CO(1-0) upper limits at $\mathrm{S/N>4}$ following our method. These objects were unaccounted for by \citealt{Jin21}. However, as stated above, our method only integrates the positive flux within a spectral window, which can result in a flux overestimation. Consequently, we still consider these four CO(1-0) measurements as upper limits, which is reflected in Table\,\ref{T:BigTable}. These sources may have fallen near but below the S/N threshold imposed by \cite{Jin21} in their blind emission line search, thus explaining why they were not selected by them. 

To test the accuracy and reliability of our method, Fig.\,\ref{F:UpLim} compares the CO(1-0) fluxes of the sources as reported by \cite{Jin21} $\mathrm{S/N>4}$ with the measurements obtained from those same sources following our method. There is a good agreement between both approaches within the error bars. The average signal-to-noise ratio of these 10 sources following our method is $\mathrm{S/N=4.5\pm1.5}$, while \cite{Jin21} measurements yields $\mathrm{S/N=4.6\pm1.0}$. This suggests that both our spatial spaxel grid and spectral window are reasonable choices to compute upper limits over known HAEs. Thus, we compute the molecular gas mass for each member of our sample based on the results published by \cite{Jin21} on the 10 dual $\mathrm{H\alpha+CO(1-0)}$ emitters and the new upper limits that we obtained for the 33 remaining HAEs in our sample. The CO(1-0) luminosity ($L'_{\mathrm{CO(1-0)}}$) can be converted into a molecular gas mass estimate in a simple way assuming a conversion factor ($\alpha_{\mathrm{CO(1-0)}}$) that traces the amount of molecular gas from optically thick virialized clouds (\citealt{Dickman86}, \citealt{Solomon87}):

\begin{figure}    
\centering
\includegraphics[width=\linewidth]{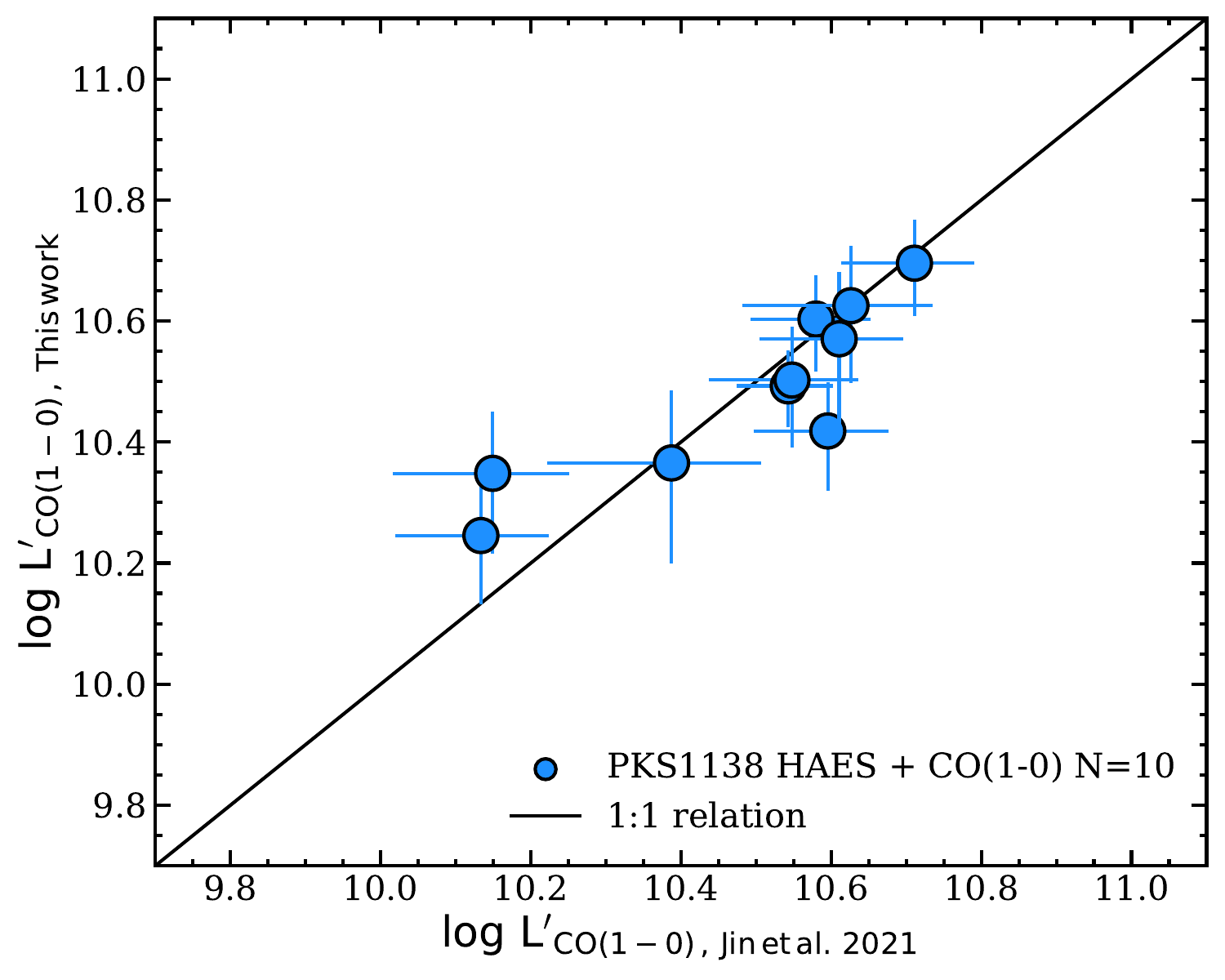}\par 
\caption{Comparison of the $L'_{\mathrm{CO(1-0)}}$ obtained by applying our methodology and the source detection codes outlined in \citet{Jin21} for HAEs with CO(1-0) detection at $\mathrm{S/N>4}$.}
\label{F:UpLim}
\end{figure}

\begin{equation}
M_{\mathrm{mol}}=\alpha_{\mathrm{CO(1-0)}}\times L'_{\mathrm{CO(1-0)}}
\label{EQ:mol}
\end{equation}
The $\alpha_{\mathrm{CO(1-0)}}$ factor is known to be approximately constant for star-forming galaxies close to the solar gas-phase metallicity value in the local Universe ($\mathrm{\alpha_{CO(1-0)}=4.36\pm0.90\,M_\odot/(K\,km\,s^{-1}\,pc^{2})}$; \citealt{Bolatto13}) albeit it depends on additional factors such as temperature, cloud density and metallicity (e.g., \citealt{Narayanan12}). For example, lower metallicity values as those expected in the early universe may yield higher conversion factor values due to the contraction of the CO-emitting surface relative to the area where the gas is H$_2$ for a fixed cloud size (\citealt{Genzel15}; \citealt{Tacconi18}.

Some sources within our sample have gas-phase metallicity estimates based on the [{N\sc{ii}}]/H$\alpha$ ratio (\citealt{PerezMartinez23}). However, the authors showed that these sources are representative of the metal-rich end of the mass-metallicity distribution and that the scatter of such relation for individual sources is $\sim0.3$\,dex, which would introduce high uncertainties in the ($\mathrm{\alpha_{CO(1-0)}}$) determination. Taking as a reference the stacking metallicity results of \cite{PerezMartinez23} for HAEs in the Spiderweb protocluster we find that $\mathrm{12+\log(O/H)=8.29}$, 8.55 and 8.64 for mass bins at $\mathrm{8.9<\log M_*/M_\odot<10.0}$, $\mathrm{10.0<\log M_*/M_\odot<10.8}$ and $\mathrm{\log M_*/M_\odot>10.8}$ respectively. If we take Eq. 2 in \cite{Tacconi18} to compute the correction factors to $\mathrm{\alpha_{CO(1-0)}}$ based on these results we find that $\mathrm{f_{corr}=3.3}$, 1.15, and 1.0 respectively. This implies that galaxies in the intermediate and massive bin could increase their molecular gas mass by $\mathrm{<0.1\,dex}$ on average, but their effect in $\mathrm{f_{gas}}$ would be negligible. The impact becomes stronger in the low-mass and low-metallicity end. Overall, the change of $\mathrm{\alpha_{CO(1-0)}}$ would increase the molecular gas mass of $\mathrm{\log M_*/M_\odot<10.0}$ sources by $\mathrm{\sim0.5\,dex}$, also increasing their gas fractions. However, $\mathrm{f_{gas}\gtrsim0.9}$ for most of these (see Fig.\,\ref{F:fgas}) thus minimizing its impact on this property. Nevertheless, we note this correction diverges rapidly at low-metallicities and it must be used with caution in this regime. We refer to \citealt{Tacconi18} and references therein for further details. Therefore, we resort to using a single constant conversion factor value, $\mathrm{\alpha_{CO(1-0)}=4.36\,M_\odot/(K\,km\,s^{-1}\,pc^{2})}$, for our entire sample.

\subsection{Star-formation activities}  
\label{SS:SFR_method}

Our sample of protocluster HAEs for which we have obtained CO(1-0) flux measurements and upper limits has deep narrow-band and spectroscopic follow-ups (e.g., \citealt{Shimakawa15, Shimakawa18b}; \citealt{PerezMartinez23}) targeting the H$\alpha$ emission line. To compute the SFR of our HAEs, we first retrieve $\mathrm{H\alpha}$ fluxes from VLT/KMOS datacubes as published in \cite{PerezMartinez23}, which contains 32 out of the 43 HAEs analyzed in this work. We note that the spectral resolution of KMOS allows us to separate the H$\alpha$ and [N{\sc{ii}}] emission lines in our sources (see \citealt{PerezMartinez23}) for further details). For the remaining 11 HAEs, we resort to the available narrow-band imaging to extract their $\mathrm{H\alpha}$ fluxes following \cite{Shimakawa18b}. We correct for filter transmission and [N{\sc{ii}}] contamination by using the mass-metallicity relation from \cite{PerezMartinez23} and our SED-based stellar masses (see Sect.\,\ref{SS:SED}). Then, we use the measurements of H$\alpha$ from these past works as a proxy for star formation within our sample and apply the \cite{Kennicutt98} calibration modified for a Chabrier IMF to compute their star-formation rates (SFR). This calibration has proven to be one of the most reliable both at local and high redshift (e.g. \citealt{Moustakas06}, \citealt{Wisnioski19}):
\begin{equation}
\mathrm{SFR(H\alpha)}=4.65\times10^{-42}\times L(\mathrm{H\alpha})
\label{EQ:SFR}
\end{equation}
where $L(H\alpha)$ is the luminosity of the H$\alpha$ emission-line. We assume a Calzetti's extinction law ($R_v$=4.05, \citealt{Calzetti2000}) to account for the dust attenuation and follow the extinction correction calibration presented in \cite{Wuyts13} to obtain our final SFR values. In this calibration, the total extinction over the H$\alpha$ line ($\mathrm{A(H\alpha)}$) is computed as a sum of both the continuum and the nebular contributions produced in active star-forming regions yielding $\mathrm{A(H\alpha)=A_{cont}+A_{extra}}$. We use the extinction $A_v$ values obtained from SED fitting to account for the diffuse dust attenuation in the galaxy's continuum following $\mathrm{A_{cont}=0.82A_{v,SED}}$, while $\mathrm{A_{extra}=0.9A_{cont}-0.15A_{cont}^{2}}$ (\citealt{Wuyts13}; \citealt{Wisnioski19}). This calibration is in good agreement with the previous extinction estimates made by \cite{Calzetti2000} in the local universe. Our sample of 43 HAEs includes 9 X-ray emitters (\citealt{Tozzi22a}), which are likely indicative of AGN activity. Two of them are identified as broad-line AGN (ID 647 and 911) by \cite{PerezMartinez23} and their SFR is computed from their decomposed narrow-line component. Nevertheless, we are unable to quantify possible contamination from AGN activity or shocks into the narrow-line component of any of our sources and assume that it originates from star-formation activities in the following sections. The SFR of our protocluster sample can be found at the end of this work in Table \ref{T:BigTable}.

\subsection{Stellar masses and dust attenuation}
\label{SS:SED}

The field of the Spiderweb protocluster is covered with extensive and deep photometry comprising the rest-frame X-ray to NIR wavelength range for galaxies at $z=2.16$ (see Sect.\,\ref{S:Data}). We make use of SED fitting over the photometry retrieved from these datasets to derive stellar masses and dust extinctions for our sample of HAEs following the same procedure outlined in \cite{PerezMartinez23} and using the same photometric bands. In the next paragraphs, we provide a brief summary of the main steps carried out during this process. First, we take the Subaru/MOIRCS narrow-band image as the reference in terms of seeing ($\mathrm{FWHM=0.7\arcsec}$) and pixel size ($\mathrm{0.167\arcsec}$). We carried out PSF matching and pixel scale resampling over the remaining imaging except for the Subaru/Suprime-Cam B-band and the Spitzer/IRAC bands whose PSF is significantly larger (i.e., $\mathrm{FWHM\approx1.1\arcsec}$ and $\mathrm{\sim2\arcsec}$ respectively), than that of the narrow-band images. 

Then, we perform dual-image photometry with \textit{SExtractor} (\citealt{Bertin96}) using the NB2071 image for source detection while measuring the observed magnitudes (MAG\_AUTO) on the rest of the images. In addition, we carry out independent single-band photometry for the Subaru/Suprime-Cam \textit{B}-band and the Spitzer/IRAC 3.6\,$\mathrm{\mu m}$ and 4.5\,$\mathrm{\mu m}$ bands, although only $\sim60\%$ of our sample is covered by the Spitzer available imaging. As a result, we construct a 12-band multiwavelength catalog for our sources encompassing the rest frame $\mathrm{1400-14200}$\,\AA\, wavelength range for protocluster members at $z\approx2.16$. Each of our sources is covered by two to five bands with central wavelengths beyond $\mathrm{4000\,\AA}$ in the rest frame (i.e., HAWKI/VLT \textit{H} and \textit{Ks} bands, Subaru/MOIRCS \textit{Ks} band, and IRAC 3.6 and $\mathrm{4.5\,\mu m}$). We run the SED fitting code CIGALE (Code Investigating GALaxy Emission, \citealt{Boquien19}) over this multiband catalog to derive the stellar mass and dust extinction for the sources within our sample. CIGALE is a widely used SED fitting routine that combines the modeling of composite stellar populations with nebular emission and dust attenuation while conserving the energy balance between the UV and the IR. 

\begin{table}
\caption{Summary of CIGALE SED fitting modules and parameter configuration.}
\label{T:CIGALE}
\centering
\begin{tabular}{lc}
\hline
\noalign{\vskip 0.1cm}
CIGALE Module & Parameter Values \\
\hline
\noalign{\vskip 0.1cm}
\textbf{Stellar Pop.} & \cite{BC03} \\
IMF & \cite{Chabrier03} \\
Metallicity   & 0.004 \\
\hline 
\noalign{\vskip 0.1cm}
\textbf{SFH}& Exp. Delayed \\
$\mathrm{\tau_{main}\,(Myr)}$ & $\mathrm{100,\,500,\,1000,\,2000,\,4000,\,6000,\,8000}$ \\
$\mathrm{age_{main}\,(Myr)}$ & $500, 1000, 1500, 2000, 2500, 3000$ \\
$\mathrm{f_{burst}}$ & 0.01, 0.05, 0.10 \\ 
$\mathrm{\tau_{burst}\,(Myr)}$ & $\mathrm{20, 50, 70, 100, 300}$ \\ 
\hline 
\noalign{\vskip 0.1cm}
\textbf{Neb. Emission}&\\
$\log U$ & -2.4, -2.5, -2.6, -2.7, -2.8\\
$\mathrm{f_{esc}}$ & 0.0 \\
\hline 
\noalign{\vskip 0.1cm}
\textbf{Dust extinction} & \cite{Calzetti2000} \\
$E(B - V)_\mathrm{lines}$ &  0 to 2 \\
$E(B - V)_\mathrm{factor}$ & 0.44 \\
$R_v$& 4.05\\
\noalign{\vskip 0.0cm}
\hline 
\end{tabular}
\end{table}

We create a grid of SED models based on CIGALE (\citealt{Burgarella05}; \citealt{Noll09}; \citealt{Boquien19}) modules and parameters. Our grid is based on the stellar population synthesis models of Bruzual \& Charlot (\citealt{BC03}) using the Chabrier IMF (\citealt{Chabrier03}) and subsolar metallicity (i.e $Z = 0.004$). We assume an exponentially delayed star formation history with possible e-folding times between 0.1 and 8 Gyr and stellar population ages constraint to be younger than the age of the Universe at $z=2.16$ (i.e., $\sim3$ Gyr). We allow for the presence of a recent ($\mathrm{<300\,Myr\,old}$) but small ($\mathrm{<10\%\,in\,mass}$) burst of star formation to account for possible recent interactions within the overdense protocluster environment. We account for the effects of nebular emission by setting the ionization parameter ($U$) range at $-2.4<\log(U)<-2.8$, which describes typical values for star-forming galaxies at $z\sim2$ (\citealt{Cullen16}). Then, we apply Calzetti’s attenuation law (\citealt{Calzetti2000}) with extinction values ranging $\mathrm{E\left( B-V\right)_s=0-2}$ mag in steps of 0.1 mag. Table\,\ref{T:CIGALE} summarizes our CIGALE configuration. The parameters not included in this table are set to their default values.

\begin{figure}
\includegraphics[width=\linewidth]{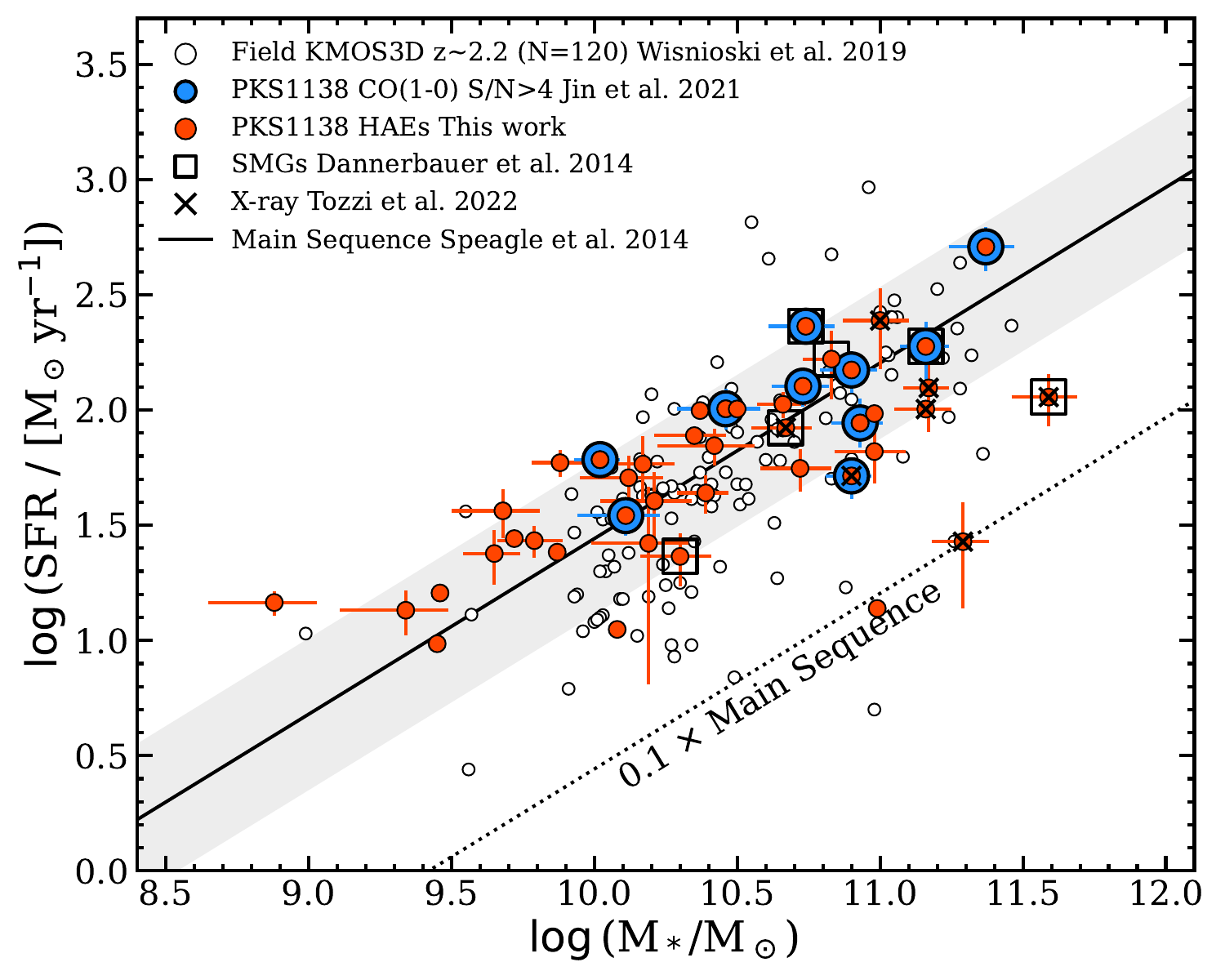}\par 
\caption{Star forming main sequence diagram. The black solid line and shaded grey region depict the Main Sequence of star formation at $\mathrm{z=2.16}$ from \citet{Speagle14}. Red small circles the spectroscopically confirmed HAEs of \citealt{Shimakawa18b} and \cite{PerezMartinez23} within the ATCA datacube footprint. Blue circles show the location of the CO(1-0) emitters at $\mathrm{S/N>4}$ from \citet{Jin21} which are part of our sample of HAEs. For context, we also add empty squares to show the overlap with the sample of SMGs in this field from \citet{Dannerbauer14} and with the X-ray emitters reported by \citet{Tozzi22a}.}

\label{F:MS}
\end{figure}

CIGALE follows a Bayesian-like approach by evaluating and weighing the models within a given grid depending on their proximity to the best fit, which has the heaviest weight. Based on the distribution of weights across the grid of models, the final physical properties and their uncertainties are estimated. This approach takes into account the uncertainties of the observations while also including the effect of intrinsic degeneracies between physical parameters (\citealt{Boquien19}). Finally, the physical properties and their uncertainties are estimated as the likelihood-weighted means and standard deviations. The stellar mass and V-band attenuation ($\mathrm{A_V}$) values obtained for our sample carry uncertainties of the order of $\mathrm{\pm0.1\,dex}$ and $\mathrm{\pm0.2\,mag}$ respectively. In Fig.\,\ref{F:MS}, we put our sample in context by comparing it with the so-called "Main Sequence" of star formation (\citealt{Speagle14}) at $\mathrm{z\approx2.16}$ similarly to what was shown in \cite{PerezMartinez23} as we share 32 sources in common out of the 43 that comprise our HAE sample. Both galaxies with CO(1-0) detections and upper limits scatter around the same locus and agree within uncertainties with the expectations from the coeval Main Sequence except for a few cases.

\subsection{Environmental proxies}
\label{SS:Environment}

In this section, we aim to characterize the environment where our sources reside by inspecting both their immediate surroundings (i.e., local density) as well as their association with the larger scale structure of the Spiderweb protocluster (i.e., phase space). However, the quantification of the environment within overdense regions at high-z remains a challenge for most observational works due to three main reasons: projection effects, sample incompleteness, and structure definition. We can mitigate the first one thanks to the large number of spectroscopic redshifts ($\mathrm{>100}$) and narrow-band detections (HAEs and LAEs) available across the Spiderweb protocluster field (see Sect.\,\ref{S:Intro}). Nonetheless, limited projection effects could still be present due to the intrinsic proper motions of galaxies within protoclusters (\citealt{Overzier16}). The second is only partially alleviated thanks to the diversity of galaxy populations studied in this field over the last two decades, confirming the membership of a large number of both dust-obscured and unobscured star-forming galaxies and AGN throughout different methods (see Sect.\,\ref{S:Intro}), albeit the existence of a passive population and its distribution remains unexplored up to now. Finally, protoclusters are by definition large-scale structures studied amid their assembly process, and as such, measurements of their current size and total mass are highly uncertain (\citealt{Muldrew15}). The discovery of a nascent ICM in the Spiderweb protocluster via Sunyaev-Zeldovich (SZ) effect confirms that this structure will evolve into a galaxy cluster by $\mathrm{z=0}$ (\citealt{DiMascolo23}) albeit the inner core might not yet be fully virialized, with current dynamical and SZ total mass estimates differing by a factor 2-4 (see \citealt{Saro09}; \citealt{Shimakawa14}; \citealt{DiMascolo23}). 
Nevertheless, we attempt to quantify the global environment by assuming the virialization of the protocluster core and inspecting the caustic profiles ($\eta$, \citealt{Haines12}; \citealt{Noble13}) in the phase-space diagram similarly to \citet{Wang18} and \citet{PerezMartinez23}: 
\begin{equation}
    \eta=(R_{\mathrm{proj}}/R_{200})\times(\left | \Delta v \right |/\sigma)
\label{EQ:GlobalDensity}
\end{equation}
where $\mathrm{R_{proj}/R_{200}}$ is the projected clustercentric distance relative to $\mathrm{R_{200}}$. This is the radius enclosing a volume with a mass density 200 times the critical density of the universe at the cluster's redshift. In our analysis, we assume the Spiderweb Galaxy as the center of this protocluster. By comparison, the centroid of the X-ray diffuse emission (\citealt{Tozzi22a}; \citealt{Lepore24}) and the Sunyaev-Zeldovich signal (\citealt{DiMascolo23}) display offsets of $\mathrm{2.3\pm0.6}$ and $\mathrm{6.2\pm1.3}$ arcsecs to the Spiderweb galaxy respectively. For context, the size of the Spiderweb complex (i.e., the radio galaxy and its surrounding interacting debris) encompasses at least 6-8 arcsecs (\citealt{Miley06}, \citealt{Hatch08}, \citealt{Kuiper11}) while our sources are distributed up to 3 arcminutes away from the Spiderweb galaxy. Thus, we conclude that such small variations in the definition of the protocluster center have no significant impact on our environmental analysis. In addition, $\Delta v$ is the line-of-sight velocity relative to the systemic velocity of the cluster defined as $\left | \Delta v \right |= \left |(z-z_{cl})\ c/(1+z_{cl})\right |$ with $\mathrm{z_{cl}=2.156}$ being the systemic redshift of the Spiderweb galaxy (\citealt{Kurk00}). Finally, $\mathrm{\sigma}$ is the dynamically estimated inner core's velocity dispersion $\mathrm{(\sigma=683\,km\,s^{-1})}$ given in \cite{Shimakawa14}.

Alternatively, we can quantify the environment as a function of the local surface density of objects. To this end, we gather a sample of 145 unique protocluster members including spectroscopically confirmed LAEs (\citealt{Pentericci00}), narrow-band selected and spectroscopically confirmed HAEs (\citealt{Koyama13}; \citealt{Shimakawa24}), and spectroscopically confirmed CO(1-0) emitters (\citealt{Jin21}). Narrow-band selected HAEs in this protocluster have been shown to have a membership success rate of $\mathrm{>90\%}$ (\citealt{PerezMartinez23}), which is in line with the contamination rate from fore-/background sources reported by \cite{Sobral13} in the field at $\mathrm{z\sim2}$ using the same technique. Based on these results, we take narrow-band selected HAEs lacking spectroscopic confirmation as protocluster members. In addition, the survey area of our selected samples approximately overlaps with the full area of the COALAS footprint shown in Fig.\,\ref{F:MAP}, thus minimizing biases caused by uneven depths or coverage across this field. After these considerations, we compute the local surface density of objects as: 
\begin{equation}
   \Sigma_N=\frac{N}{\pi R^2_{N-1}}
\label{EQ:LocalDensity}
\end{equation}
where N is the number of sources enclosed within a radius $\mathrm{R_{N-1}}$, which is the distance to the Nth-1 nearest neighbor (i.e., not counting the galaxy of origin). The distance between sources is measured by their sky angular separation and transformed to the physical scale using a fixed cosmology described in Sect.\,\ref{S:Intro}. In this work, we compute local densities enclosing two, five, and ten neighboring galaxies (i.e. $\Sigma_2$, $\Sigma_5$, $\Sigma_{10}$). This choice allows us to trace the properties of galaxies residing in local density peaks such as close companions or pairs ($\Sigma_2$), compact groups ($\Sigma_5$), and larger overdensities ($\Sigma_{10}$).

\begin{figure*}    
\centering
\begin{multicols}{2}
\includegraphics[width=\linewidth]{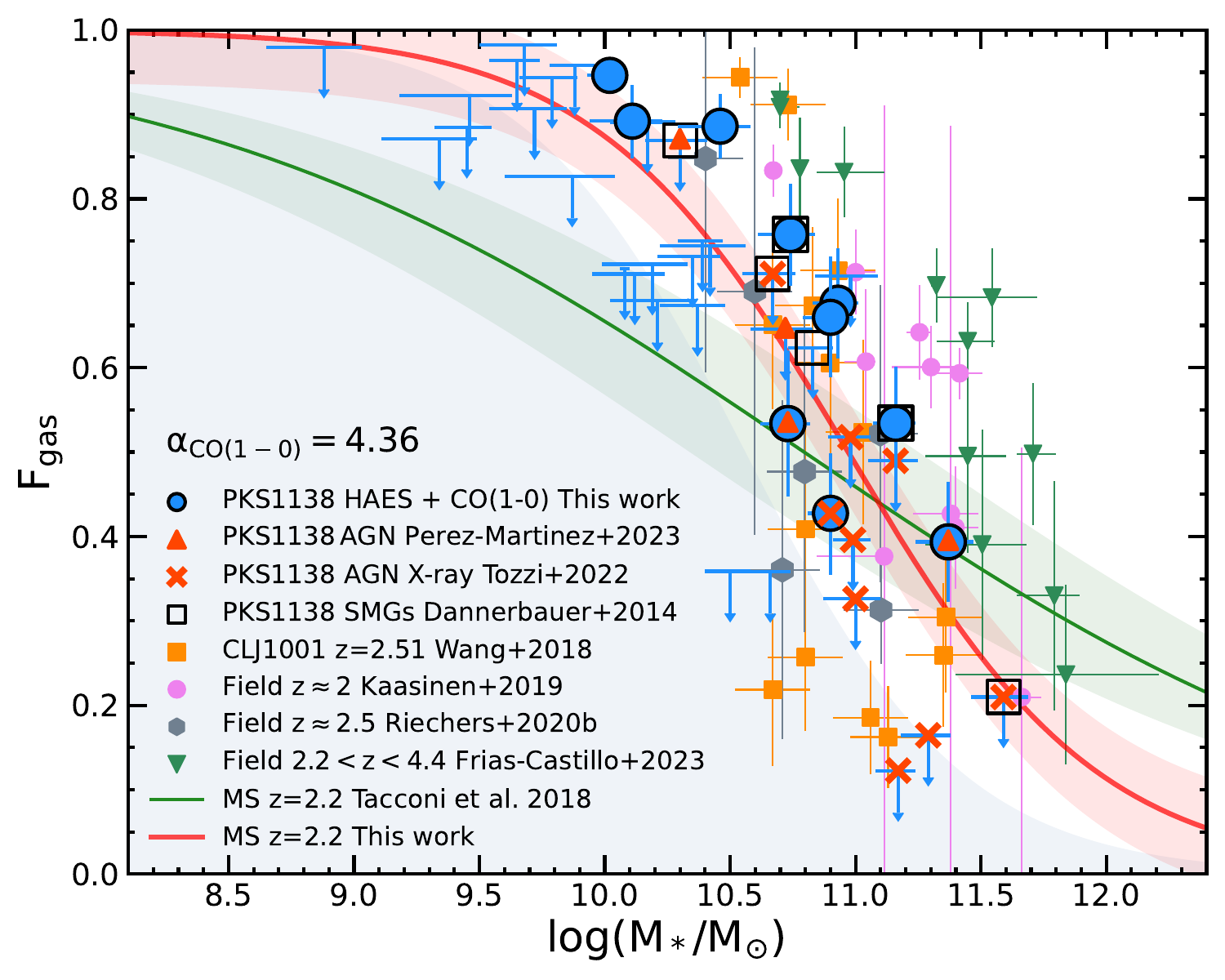}\par
\includegraphics[width=\linewidth]{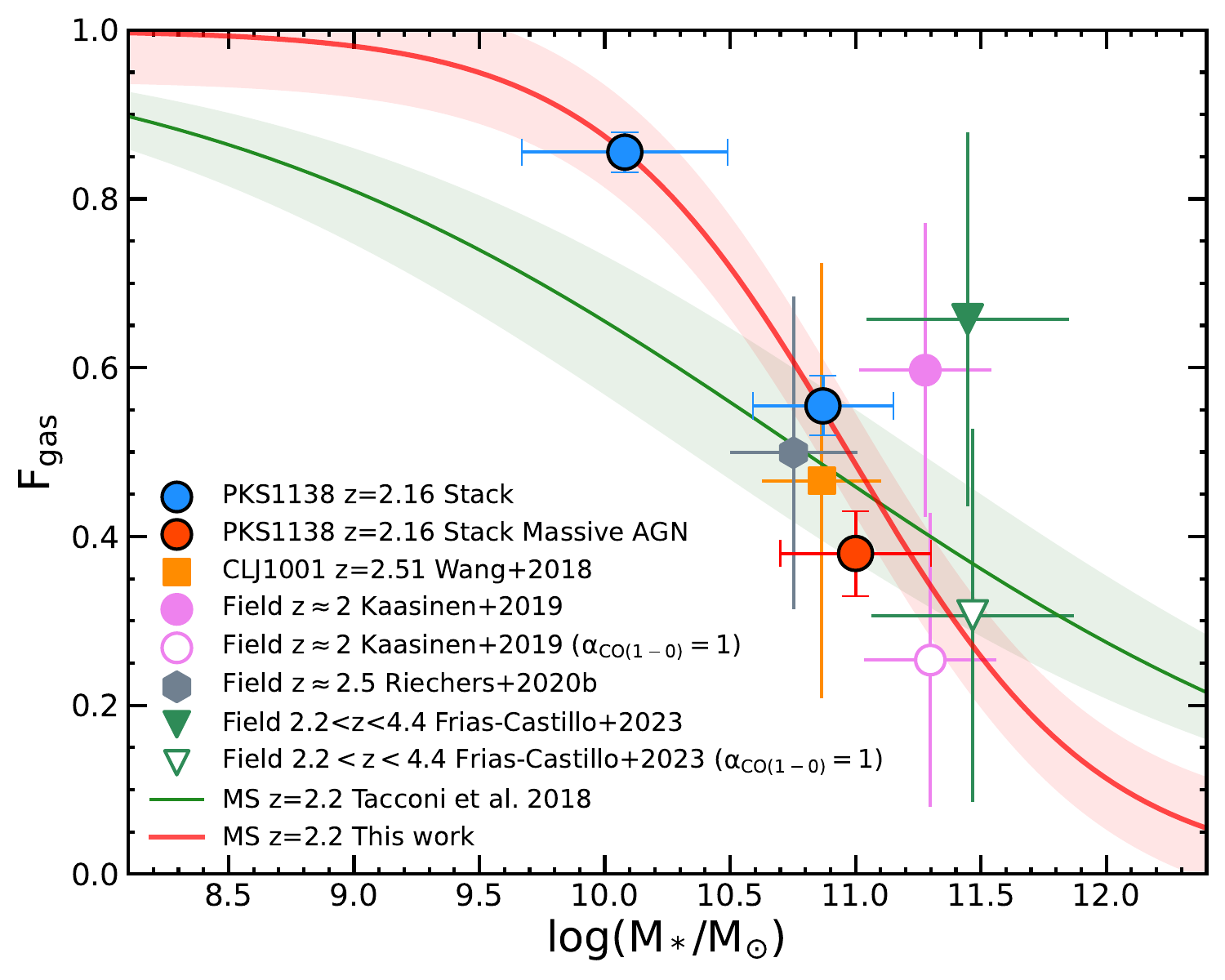}\par
\end{multicols}
    \caption{Left: Total gas fraction versus stellar mass diagram. The solid green line and shaded region represent the field scaling relations for Main Sequence (MS) galaxies from \citet{Tacconi18}. We overplot our sample in blue color (both for measurements and upper limits) and compare it with other protocluster samples at the cosmic noon such as \citet{Wang18} in orange, and coeval field samples (\citealt{Kaasinen19} in pink; \citealt{Riechers20b} in grey; \citealt{FriasCastillo23} in green). In addition, we marked those galaxies with signs of AGN activity with red crosses (X-ray emission, \citealt{Tozzi22a}) and red triangles (high [N{\sc{ii}}]/$\mathrm{H\alpha}$, \citealt{PerezMartinez23}) and those identified as SMGs by \citet{Dannerbauer14} as black empty squares. The solid red line and red area depict the fit and uncertainty to the PKS1138 and CLJ1001 protocluster samples using a logistic function, as proposed by \citet{Popping12}. The grey shaded area displays the $\mathrm{1\sigma}$ detection limit given by the median rms of the ATCA mosaic. Right: Total gas fraction versus stellar mass diagram after stacking. Our sample is divided into three bins: Low-mass galaxies ($\mathrm{\log M_*/M_\odot<10.5}$); massive galaxies ($\mathrm{\log M_*/M_\odot>10.5}$) excluding AGN candidates; and massive AGN candidates. In addition, mdian values and standard deviations for the comparison samples are displayed using the same symbol and color scheme. The empty symbols depict the median values of the \cite{Kaasinen19} and \cite{FriasCastillo23} field samples after applying $\mathrm{\alpha_{CO}\approx1}$, typical of high-z starbursts and SMGs (see Sect.\,\ref{SS:GASF} and \ref{SS:GAS_STACK}). We apply a 0.02 dex cosmetic shift in $\mathrm{M_*}$ to these two data points to improve the visibility of their error bars.}
    \label{F:fgas}
\end{figure*}

\section{Results}
\label{S:Results}

In this section, we present the molecular gas properties of our sample of 43 HAEs for which we computed CO(1-0) estimates within the Spiderweb protocluster. We recover 10 sources from \cite{Jin21} with CO(1-0) detections at $\mathrm{S/N>4}$ which represents $\mathrm{\sim23\pm12\,\%}$ of our spectroscopic sample and $\mathrm{\sim20\pm11\,\%}$ of all the detected narrow-band HAEs within the ATCA footprint (see Fig.\,\ref{F:MAP} and references therein) assuming poissonian errors. This is consistent with the CO(1-0) detection ratio found by \cite{Tadaki14} over HAEs within the USS1558-003 protocluster at $z=2.53$ at a similar depth to ours (i.e., $\mathrm{0.13-0.29\,mJy}$ per beam in $\mathrm{40\,km\,s^{-1}}$ channels). However, they could only confirm 3 detections within 20 HAEs in the relatively small area surveyed by the VLA ($\mathrm{\sim1.5\,arcmin^2}$) in that protocluster. In our case, we also obtain CO(1-0) upper limits for the remaining 33 spectroscopically confirmed HAEs. This represents the largest sample of HAEs with available CO(1-0) measurements within a single cluster in formation at $z>2$ up to date. To illustrate this further and to provide a broader context for the achievements of the COALAS project, we provide a collection of published CO(1-0) surveys carried out in clusters and protoclusters at the cosmic noon and several coeval field samples in the appendix (Table\,\ref{T:Littable}). In the following sections, we will discuss the influence of the protocluster environment over the galaxies' molecular gas reservoir and search for signs of accelerated galaxy evolution in overdense environments.

\subsection{Gas fraction of protocluster HAEs}
\label{SS:GASF}
In Fig.\,\ref{F:fgas}, we inspect the gas fraction (i.e., $\mathrm{f_{gas}=M_{mol}/(M_*+M_{mol})}$) of our sources as a function of stellar mass and in comparison with other CO(1-0) coeval protocluster (\citealt{Wang18}) and field samples (\citealt{Kaasinen19}; \citealt{Riechers20a}; \citealt{FriasCastillo23}) as well as field scaling relations (\citealt{Popping12}; \citealt{Sargent14}; \citealt{Tacconi18}). We also present our median $\mathrm{1\sigma}$ $\mathrm{F_{gas}}$ detection limit across the COALAS datacube as a function of stellar mass with a shaded grey area in Fig.\,\ref{F:fgas}. We note, however, that the rms across the ATCA mosaic is quite heterogeneous (see Sect.\,\ref{SS:COALAS} and \citealt{Jin21}). 
Consequently, our upper limits, which in some instances correspond to 1$\sigma$ measurements (Sect.,\ref{SS:CO}), naturally scatter around the boundary of this region, with variations depending on their position and associated local rms within the datacube. This constrains our ability to examine potential gas-depleted systems at the low stellar mass range (i.e., $\mathrm{\log M_*/M_\odot\lesssim10.0}$), albeit it is sufficient to investigate the process of gas depletion at the massive end.

In our sample, HAEs above and below (upper limits) $\mathrm{S/N=4}$ in CO(1-0) display a similar behavior across the entire stellar mass range. In particular, at low and intermediate stellar masses (i.e., $\mathrm{\log M_*/M_\odot<10.5}$) we detect galaxies in CO(1-0) that display gas fractions close to unity ($\mathrm{F_{gas}\gtrsim0.8}$), something we do not see for galaxies with higher stellar mass (i.e., $\mathrm{\log M_*/M_\odot\geq11.0}$). This suggests that these galaxies are still relatively young and will form the bulk of their stellar mass from their existing molecular gas reservoir. On the other hand, the most massive end displays a significant fraction of galaxies with rather low gas fraction values ($\mathrm{f_{gas}\lesssim0.4}$), indicating that if new fresh gas accretion is unable to replenish their reservoirs, they may soon start their quenching process due to the lack of cold gas to keep forming stars. Although our individual CO(1-0) data points cannot trace the lowest stellar mass regime, this trend nevertheless qualitatively agrees with the predictions of scaling relations in the field and galaxy evolution simulations.

The transition between these two regimes happens within a very narrow stellar mass range ($\mathrm{\log M_*/M_\odot=10.5-11.0}$) suggesting the presence of physical processes able to deplete the cold molecular gas reservoir of galaxies in a relatively efficient and quick way. This behavior is also reproduced by the $\mathrm{z=2.5}$ protocluster sample CLJ1001 of \cite{Wang18} and other cosmic noon field samples (\citealt{Kaasinen19}; \citealt{Riechers20a}; \citealt{FriasCastillo23}) after renormalizing their $\mathrm{M_{gas}}$ values to the conversion factor used in this work ($\mathrm{\alpha_{CO}=4.36\,M_\odot/(K\,km\,s^{-1}\,pc^{2})}$). This suggests that the physical origin of this trend might not depend on the environment. It is worth noting however that while the \cite{Riechers20b} sample overlaps well with our work and that of \cite{Wang18} at $\mathrm{2<z<3}$, the field samples of \cite{Kaasinen19} and \cite{FriasCastillo23} appear to follow parallel sequences offset towards higher stellar masses. This offset is likely caused by sample selection effects as these two studies are predominately focused on massive ($\mathrm{>2\times10^{10}\,M_\odot}$) IR-bright sources (e.g., SMGs selected by Herschel or ALMA), while our work traces stellar masses down to $\mathrm{\gtrsim10^{9}\,M_\odot}$ and our parent sample is dominated by typical main sequence HAEs at the cosmic noon (see Fig.\,\ref{EQ:SFR}, but also \citealt{Shimakawa18b}; \citealt{PerezMartinez23}). This also applies to the field sample of \cite{Riechers20a}, which largely overlaps with the expected properties of Main Sequence galaxies as shown by \cite{Aravena19}. We note that if instead we apply a conversion factor typical of high-z SMGs (i.e., $\mathrm{\alpha_{CO}\approx1\,M_\odot/(K\,km\,s^{-1}\,pc^{2})}$, \citealt{Hodge12}; \citealt{CalistroRivera18}; \citealt{Birkin21}; \citealt{Riechers21}; \citealt{FriasCastillo23}; \citealt{Liao23}) the median gas fraction of these two field samples would drop to $\mathrm{f_{gas}\approx0.25-0.30}$, consistent with the scatter of our sample at the massive end. We present some of the basic properties of these CO(1-0) samples in Table\,\ref{T:Littable}, thus exposing additional differences in terms of median $\mathrm{L'_{CO(1-0)}}$ and surveyed area. Finally, we have excluded from our comparisons other literature protocluster samples whose gas masses are obtained using higher CO transitions (e.g., \citealt{Tadaki19}) or dust continuum (e.g., \citealt{Zavala19}; \citealt{Aoyama22}) to prevent the propagation of additional uncertainties associated with their gas conversions. Nevertheless, these samples qualitatively agree with our trend albeit with a larger scatter. 

Interestingly, the sharp change of properties at $\mathrm{\log M_*/M_\odot=10.5-11.0}$ seen in our sample was already reported by \cite{Popping12} who attempted to model it by fitting a logistic function with the form:
\begin{equation}
\frac{\mathrm{M_{mol}}}{\mathrm{M_{mol}+M_*}}=\frac{1}{\mathrm{\exp{(-B\times(\log M_*-A))}+1}}
\label{EQ:logistic}
\end{equation}
where A and B are fitting parameters that depend on redshift in \cite{Popping12}. This function is shown in Fig.\,\ref{F:fgas} by a dashed black line. Given that our sample displays a similar behavior, we use this same functional form to fit our protocluster sample and that of \cite{Wang18} simultaneously obtaining that $\mathrm{A=10.93\pm0.05}$ and $\mathrm{B=-2.15\pm0.33}$ (red solid line and shaded red area in Fig.\,\ref{F:fgas}). Removing \cite{Wang18} from the fit would slightly smoothen the fit, thus increasing the values of A and B by $1\sigma$ and $0.5\sigma$ sigma respectively. Furthermore, based on previous works, we identified additional subpopulations such as submillimeter galaxies (SMGs, \citealt{Dannerbauer14}) and X-ray emitters (\citealt{Tozzi22a}) and emission line diagnostic AGN candidates (\citealt{PerezMartinez23}) within our sample of 43 HAEs. Intriguingly, we find that while the SMGs are distributed across the whole range covered by our sample in the $\mathrm{f_{gas}-M_*}$ plane, the AGN candidates appear only at $\mathrm{\log M_*/M_\odot\geq10.5}$ and dominate the low gas fraction regime in our sample with 5 out of 6 sources at $\mathrm{f_{gas}\leq0.4}$. \cite{Tozzi22a} reported the AGN fraction is six times above the coeval field in this protocluster based on their X-ray Chandra imaging. This enhancement is driven however by relatively X-ray bright ($\mathrm{L_{2-10keV}>4\times10^{43}\,erg\,s^{-1}}$) and massive ($\mathrm{\log M_*/M_\odot\geq10.5}$) galaxies in agreement with our findings. Nevertheless, \cite{Tozzi22a} only find marginal detection corresponding to luminosities $\mathrm{<10^{41}\,erg\,s^{-1}}$ when stacking both the spectroscopically confirmed and narrow-band selected protocluster members in the X-ray images below these thresholds, which indicates no significant AGN activity at lower masses and X-ray luminosities.

\subsection{Gas fraction: Stacking analysis}
\label{SS:GAS_STACK}

\begin{table}
\centering
\caption{Number of sources, median stellar masses, and gas fractions derived after the stacking analysis for the three bins of our sample: HAEs with $\mathrm{M_*<10^{10.5}\,M_\odot}$; HAEs with $\mathrm{M_*>10^{10.5}\,M_\odot}$; and AGN candidates, all of which also display stellar masses above $\mathrm{10^{10.5}\,M_\odot}$.}
\begin{tabular}{cccc}
\hline
\noalign{\vskip 0.1cm}
ID & N  & $\mathrm{\log\,M_*/M_\odot}$  & $\mathrm{F_{gas}}$  \\
\noalign{\vskip 0.1cm}
\hline
\noalign{\vskip 0.1cm}
Low-mass & 23 & $\mathrm{10.08_{-0.07}^{+0.06}}$ & $\mathrm{0.86\pm0.04}$ \\ 
\noalign{\vskip 0.1cm}
High-mass (No AGN) & 8 & $\mathrm{10.87_{-0.05}^{+0.04}}$ & $\mathrm{0.55\pm0.04}$ \\
\noalign{\vskip 0.1cm}
High mass (AGN only) & 12 & $\mathrm{11.00_{-0.10}^{+0.08}}$ & $\mathrm{0.38\pm0.05}$  \\
\noalign{\vskip 0.1cm}
\hline 
\noalign{\vskip 0.0cm}
\noalign{\vskip 0.0cm}
\end{tabular}
\label{T:Stack}
\end{table}

\begin{figure*}    
\centering
\begin{multicols}{3}
\includegraphics[width=\linewidth]{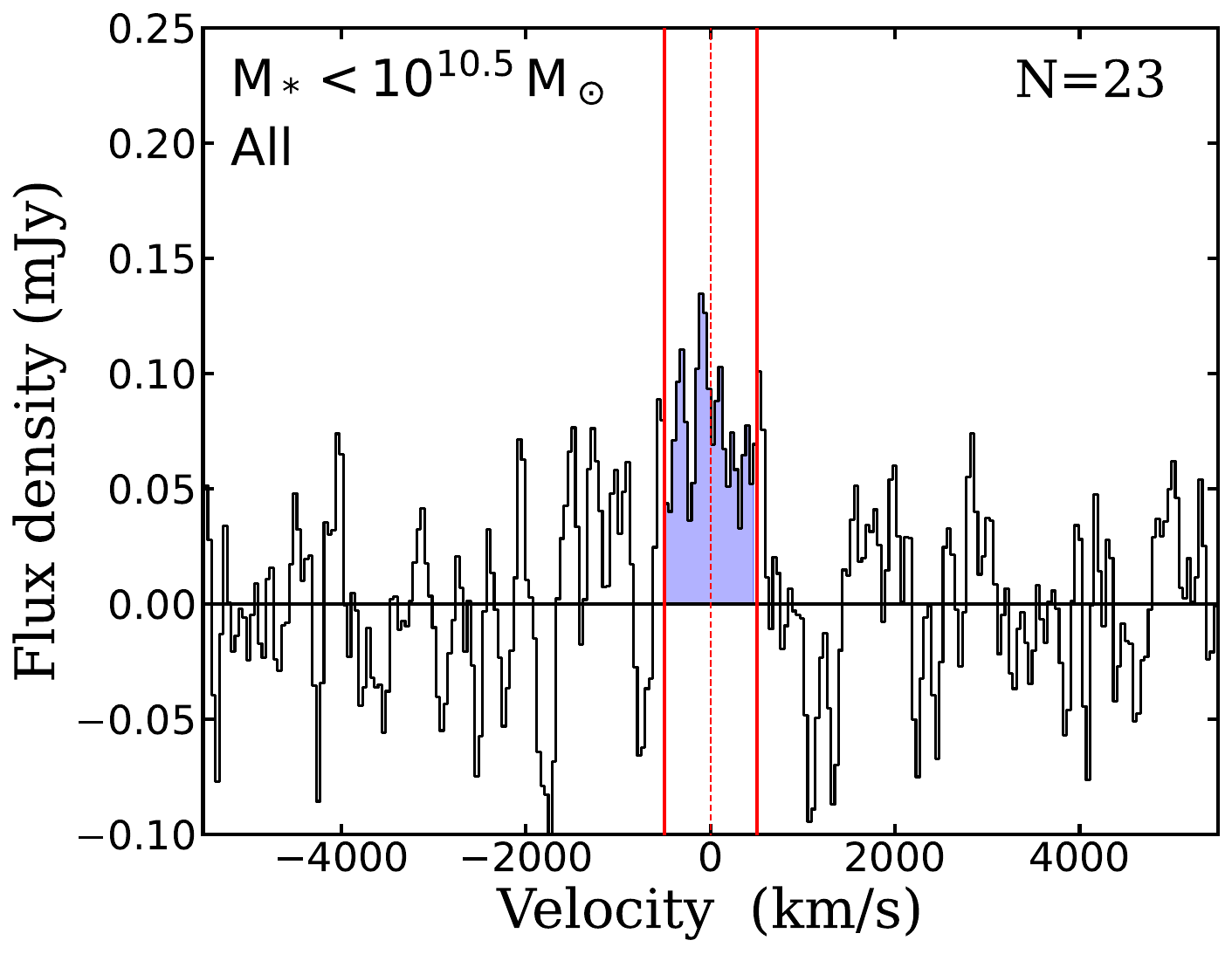}\par
\includegraphics[width=\linewidth]{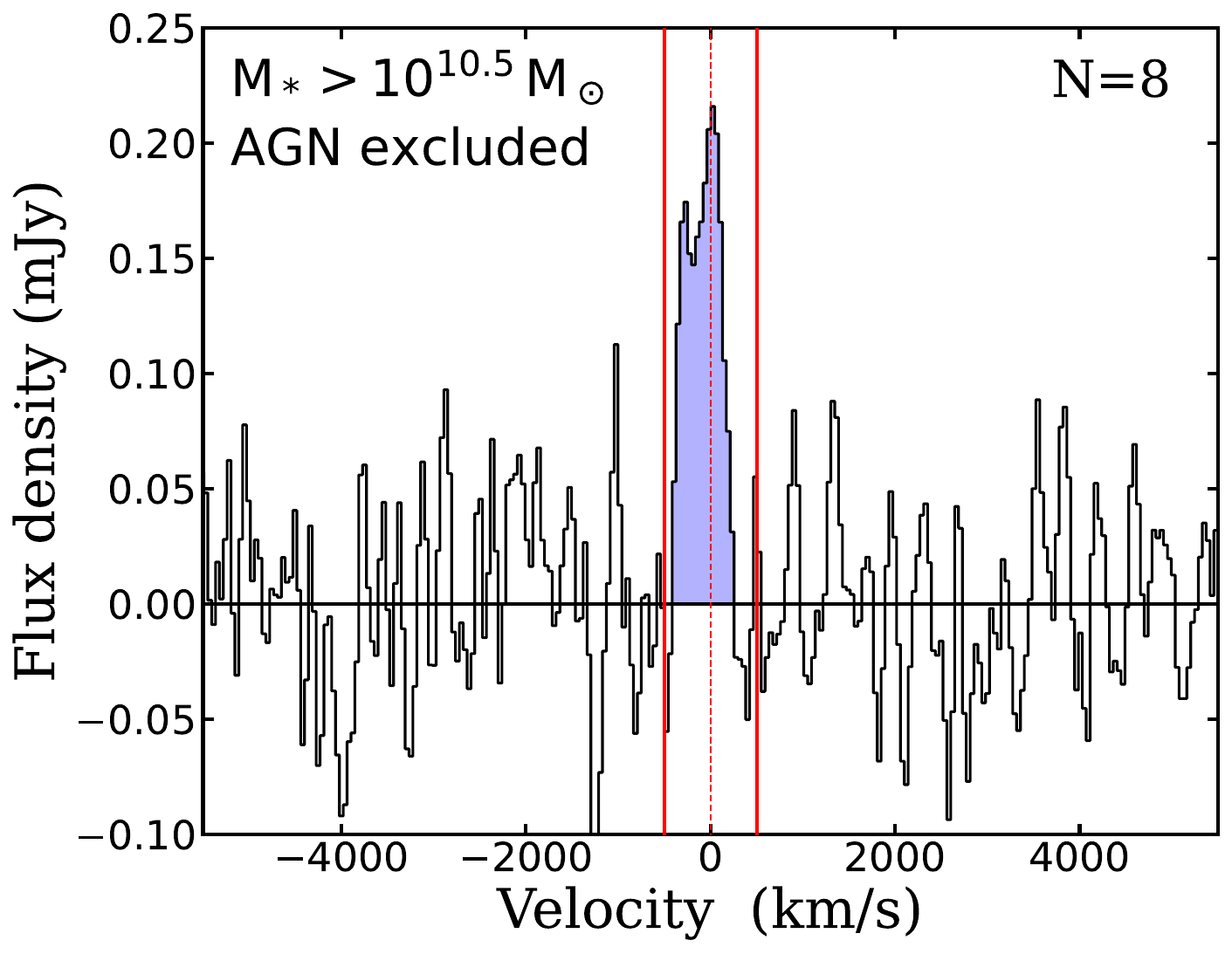}\par
\includegraphics[width=\linewidth]{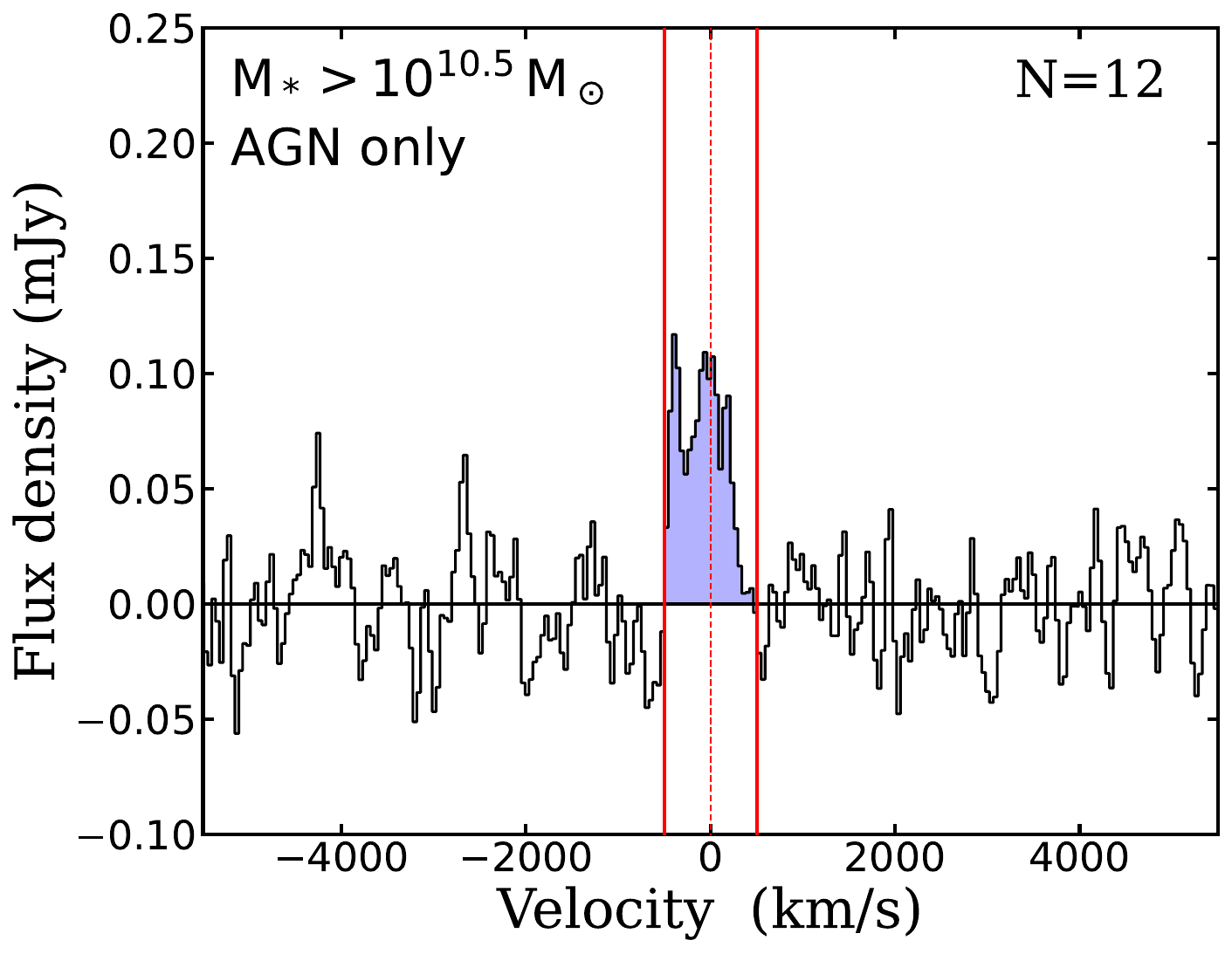}\par
\end{multicols}
    \caption{Stacked spectra around the CO(1-0) emission line for the three bins defined in Sect.\,\ref{SS:GAS_STACK}: Low-mass galaxies ($\mathrm{\log M_*/M_\odot<10.5}$), massive galaxies ($\mathrm{\log M_*/M_\odot>10.5}$) excluding AGN candidates, and massive AGN candidates.}
        \label{F:Fstack}
\end{figure*}

To gain further insights into the typical gas fraction of protocluster HAEs across different mass regimes, we divide our sample into two bins and resort to stacking analysis regarding their CO(1-0) emission. Low-mass HAEs encompass 23 objects with $\mathrm{\log\,M_*/M_\odot<10.5\,M_\odot}$ while another 20 sources display $\mathrm{\log\,M_*/M_\odot>10.5\,M_\odot}$. This latter group is divided in two bins, one representing massive HAEs without signs of AGN activity contains 8 galaxies, and another for the remaining 12 massive sources that are considered AGN candidates based on their X-ray emission or line diagnostics. The stacking analysis is performed using a similar approach than in \cite{PerezMartinez24}. We stack the spectra of sources within each bin weighted by their individual noise values: 
\begin{equation}
\mathrm{F_{stack}}=\sum_{i}^{n}\frac{F_{i}(v)}{\sigma_{i}^2}\Bigg/\sum_{i}^{n} \frac{1}{\sigma_{i}^2}
\label{EQ:stack}
\end{equation}
where $F_{i}(v)$ is the peak flux density as a function of velocity for a given source, and $\sigma_{i}$ is the average noise as described by \cite{Jin21} for the location of each source in the COALAS datacube. We use the spectroscopic redshift of our sources to shift their expected CO(1-0) emission to $\mathrm{v=0\,km/s}$ before carrying out the stacking procedure. Fig.\,\ref{F:Fstack} displays the resulting stacked spectrum for each bin defined above. We measure the CO(1-0) line flux following the same procedure described in Sect.\,\ref{SS:CO} for our individual sources, and compute its uncertainty by measuring the standard deviation of two regions $\mathrm{2500\,\AA}$ wide to the left and right of the measured emission line velocity window (red solid lines in Fig.\,\ref{F:Fstack}) and separated by $\mathrm{\pm500\,km/s}$ from it.

The results of this process are displayed in Table\,\ref{T:Stack} and in the right-hand panel of Fig.\,\ref{F:fgas} (blue and red circles). In addition, we show the median $\mathrm{F_{gas}}$ and $\mathrm{M_*}$ for the rest of protocluster and field comparison samples assuming $\mathrm{\alpha_{CO(1-0)}=4.36}$ (filled symbols). For reference, and as discussed in Sect.\,\ref{SS:GASF}, we compute the median values of the field samples of \cite{Kaasinen19} and \cite{FriasCastillo23} when assuming $\mathrm{\alpha_{CO(1-0)}=1}$ (empty pink circle and green triangle). This is a more appropriate conversion factor for high-z starbursts and SMGs (e.g., \citealt{Hodge12}) and would shift these two samples from a parallel sequence to Eq.\,\ref{EQ:logistic} as described in Sect.\,\ref{SS:GASF} to the massive and gas depleted end of our sequence (red solid line and shaded area in Fig.\,\ref{F:fgas}). In such case and given the high SFR of those samples, they may deplete their gas reservoir and become passive within a very short period. The three bins of our sample define a sequence where galaxies at low and intermediate stellar mass display high gas fractions ($\mathrm{F_{gas}\approx0.86\pm0.0.04}$). However, the massive end display diminished gas fraction values, with $\mathrm{F_{gas}\approx0.55\pm0.04}$ for massive HAEs without signs of AGN activity and $\mathrm{F_{gas}\approx0.38\pm0.05}$ for massive HAEs labeled as AGN candidates. This supports the scenario where both stellar-mass growth and AGN activity contribute to the gradual exhaustion of the gas reservoir in our sample, particularly for objects with a stellar mass beyond  $\mathrm{\log\,M_*/M_\odot>10.5\,M_\odot}$.

\subsection{Depletion times}

\begin{figure}
\includegraphics[width=\linewidth]{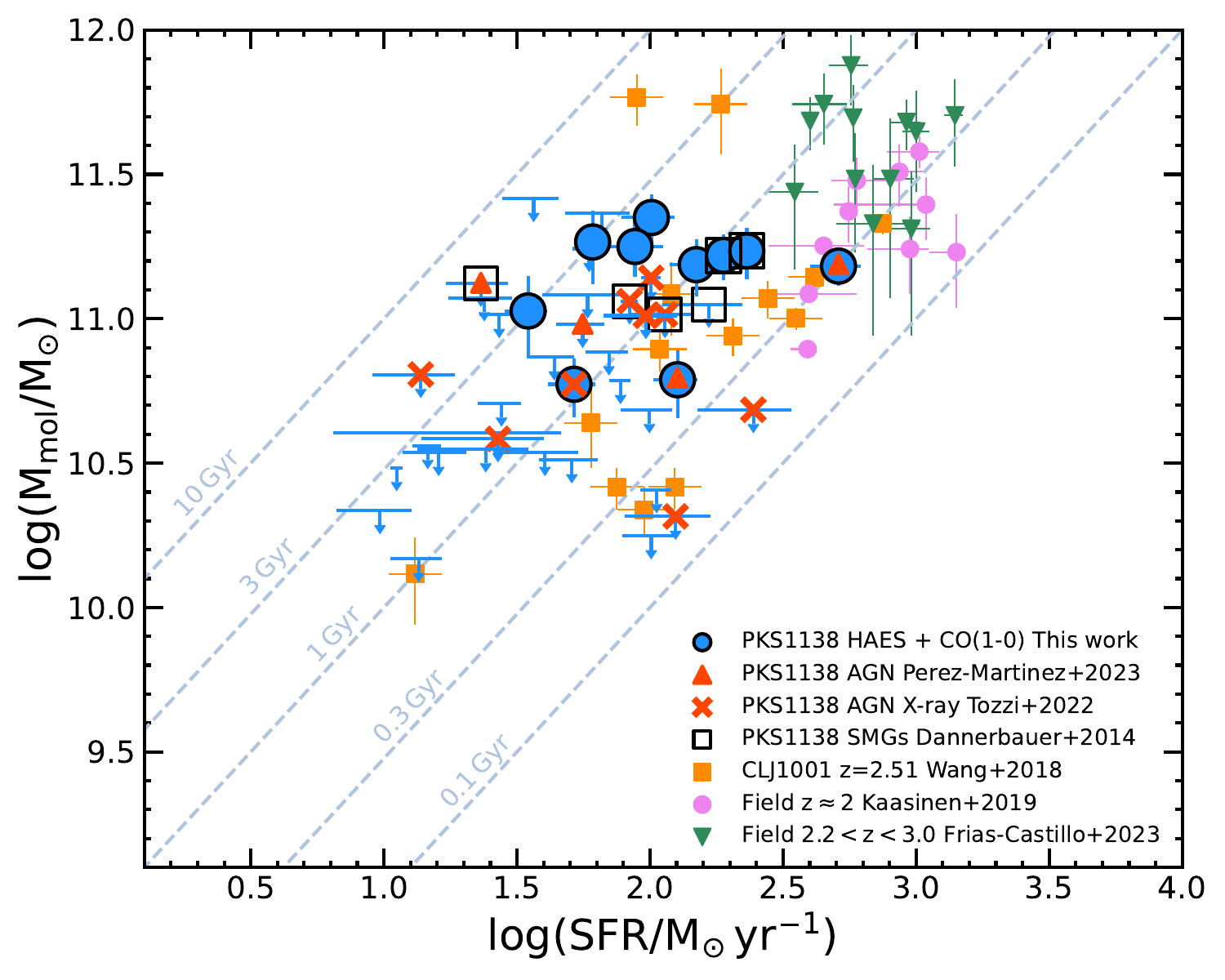}\par 
\caption{Molecular gas mass versus star formation rate diagram. Colors and symbols are the same as in Fig\,\ref{F:fgas}. Dashed grey lines display depletion times between 0.1 and 10 Gyrs.}
\label{F:tdep}
\end{figure}

Similarly, we inspect the depletion times ($\mathrm{\tau_{dep}=M_{mol}/SFR}$) of our sample in Fig.\,\ref{F:tdep} using dashed gray lines over the $\mathrm{M_{mol}-SFR}$ plane. Most of our sources lie within the lines of $\mathrm{\tau_{dep}=1-3\,Gyrs}$ indicating that, in the absence of further in-/outflows of cold gas, the HAEs populating the Spiderweb protocluster at $\mathrm{z=2.16}$ will deplete their gas reservoirs and thus shut down their star formation by $\mathrm{z=1.0-1.6}$ concurring with the surge and dominance of the red sequence in the cores of the most massive galaxy clusters known at this cosmic epoch (e.g., XMMU J2235-2557 at $\mathrm{z\approx1.4}$, \citealt{Rosati09}; but also see \citealt{Grutzbauch12}; \citealt{Nantais16}; \citealt{Beifiori17}). A few objects within our sample display even lower depletion times $\mathrm{\tau_{dep}\leq0.5\,Gyrs}$ with SFRs exceeding $\mathrm{100\,M_\odot/yr}$. These objects are possibly experiencing their last episode of star formation while rapidly consuming their gas reservoirs. Furthermore, three out of the four sources within this category are labeled as AGN candidates, hinting at the co-evolution of star formation and supermassive black hole growth. 

Nevertheless, the current ICM conditions within the Spiderweb protocluster (\citealt{Tozzi22b}; \citealt{DiMascolo23}) likely still allow the channeling of cold gas streams to some of its members albeit with diminishing efficiency over time as the protocluster virializes, thus extending the lifespan of their star formation activities beyond the reported depletion times. At the same time, additional processes such as mergers (\citealt{Mei23}), gas stripping (\citealt{Boselli22}), or AGN feedback (\citealt{Heckman14}; \citealt{Dave20}) may contribute to either extending or shortening these depletion times. We will discuss the possible interpretations of these results in Sect.\,\ref{S:Discussion} and their implications in the context of galaxy evolution in overdense environments at the cosmic noon.

\subsection{Molecular gas properties and environmental effects}
\label{SS:MolEnv}
\begin{figure*}    
\centering
\begin{multicols}{2}
\includegraphics[width=\linewidth]{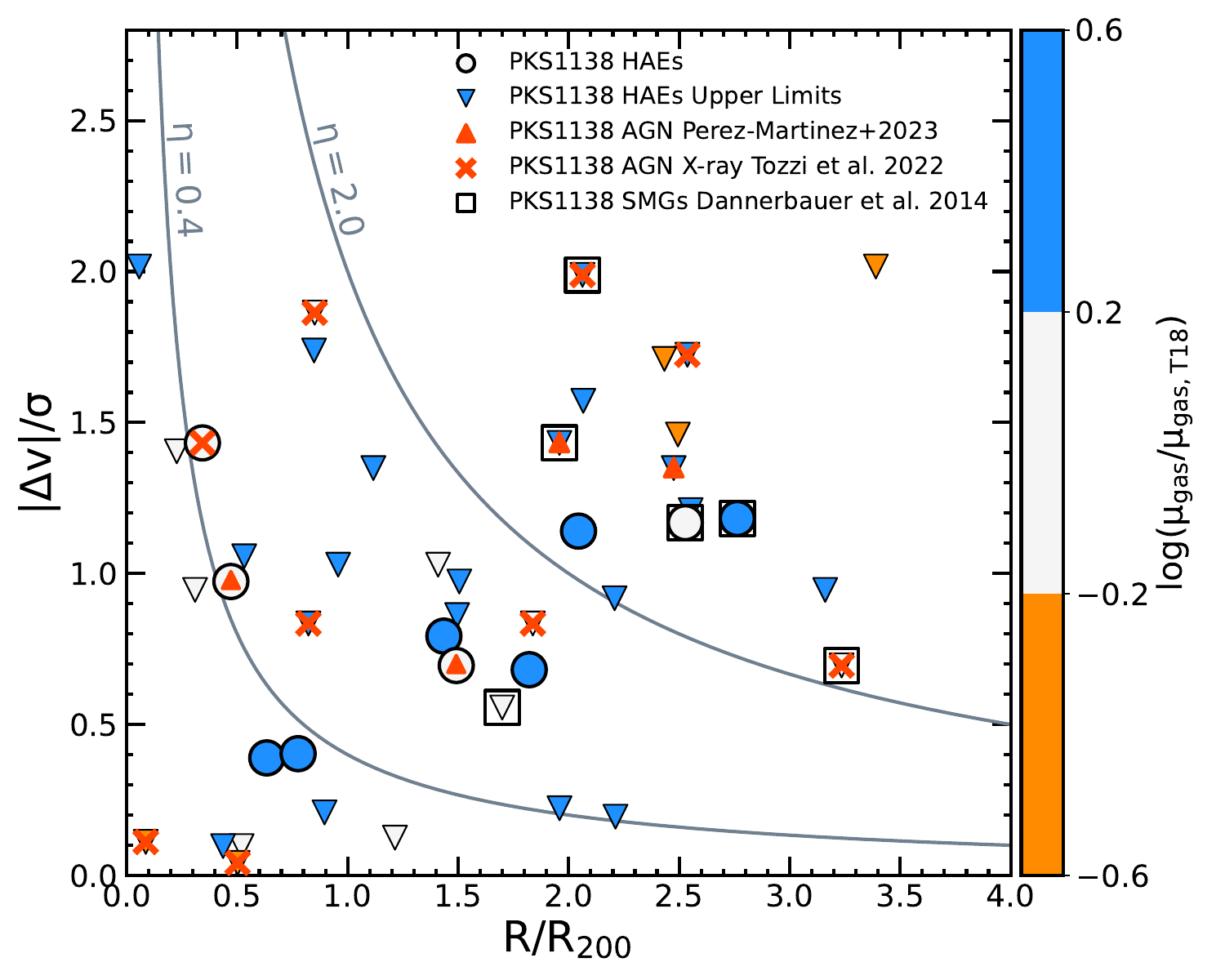}\par
\includegraphics[width=\linewidth]{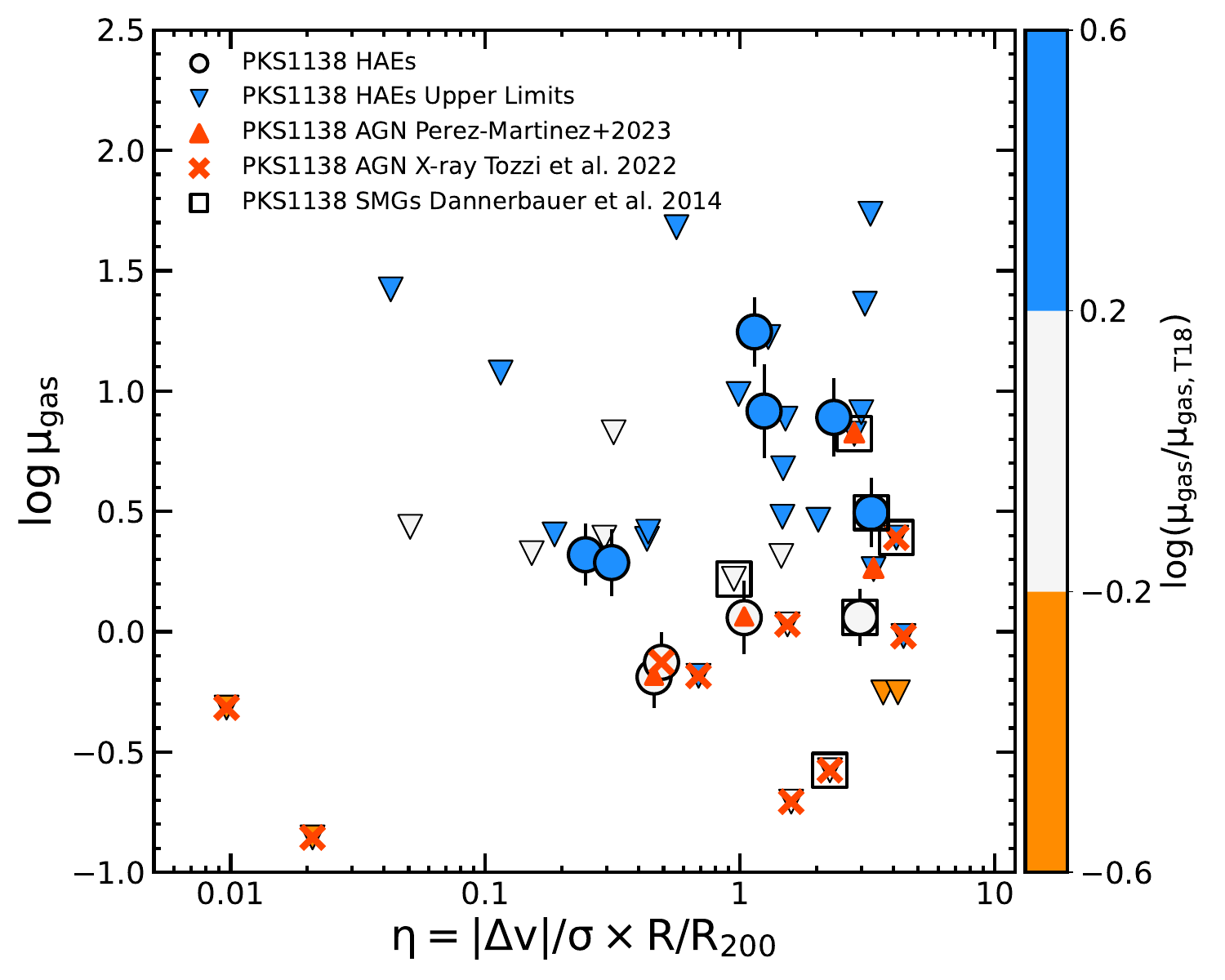}\par
\end{multicols}
\caption{Left: Phase space diagram. Objects are color-coded by their molecular to stellar mass ratio $\mathrm{\mu_{gas}}$ offsets with respect to the scaling relation of \citet{Tacconi18}. Solid grey lines divide the phase space into three regimes with galaxies being in the outskirts ($\mathrm{\eta\geq2}$), infalling ($\mathrm{0.4\leq\eta\leq2}$) or central regions ($\mathrm{\eta\leq0.4}$) similarly to \citet{PerezMartinez23}. CO(1-0) measurements at $\mathrm{S/N>4}$ are depicted as solid circles while upper limits are shown by inverted triangles. Right: Molecular to stellar mass ratio ($\mathrm{\mu_{gas}}$) offsets from the scaling relation as a function of the environmental parameter $\mathrm{\eta}$. The $\mathrm{\mu_{gas}}$ errors of the 11 CO(1-0) detections are of the order of 0.1-0.2 dex. The symbols and colors remain the same as in the left-hand diagram.}
\label{F:MolEnv}
\end{figure*}

In Fig.\,\ref{F:MolEnv} we examine the relation of the molecular-to-stellar mass ratio (i.e., $\mathrm{\mu_{gas}=M_{mol}/M_*}$) with the star formation activities within protocluster galaxies (i.e., SFR and SFE) and with the environment ($\eta$) as defined by Eq.\,\ref{EQ:GlobalDensity} (see also \citealt{Haines12}; \citealt{Noble13}). First, we show the phase-space distribution of our targets (left panel in Fig.\,\ref{F:MolEnv}) assuming the virialization of its core (\citealt{Shimakawa14}). This diagram includes both the solid $\mathrm{S/N>4}$ detection of CO(1-0) over HAEs (N=10) and the CO(1-0) upper limits following the method of Sect.\,\ref{SS:CO}. Both samples have been color-coded to reflect their offsets with respect to the $\mathrm{\mu_{gas}}$ scaling relations of \cite{Tacconi18}. In addition, we also divide the phase-space into three areas to separate galaxies in the outskirts ($\mathrm{\eta\geq2}$), infalling ($\mathrm{0.4\leq\eta\leq2}$) or central regions ($\mathrm{\eta\leq0.4}$). Most galaxies in the outskirts display $\mathrm{\mu_{gas}}$ values more than 0.2 dex above the scaling relation of \cite{Tacconi18}, while a lower number of them can be found in the infalling and core regions, which hints at the presence of some weak environmental segregation. In the right-hand panel, we find a similar picture over the $\mathrm{\log\mu_{gas}-\eta}$ plane with galaxies with higher absolute $\mathrm{\mu_{gas}}$ values populating the outskirts of the protocluster (i.e., high $\mathrm{\eta}$) and a declining trend in $\mathrm{\mu_{gas}}$ towards the cluster core (i.e., low $\mathrm{\eta}$). Similar trends were also reported by \cite{Wang18} in CLJ\,1001+0220 at $\mathrm{z=2.506}$ as a function of $\eta$, albeit most of their sources are concentrated in the protocluster inner region ($\mathrm{R<150\,kpc}$) and their total surveyed area is relatively small (5.3\,$\mathrm{arcmin^2}$) while this work examines a large mosaic of 25.8\,$\mathrm{arcmin^2}$ thus probing a larger range of environments. 

\begin{figure*}    
\centering
\begin{multicols}{2}
\includegraphics[width=\linewidth]{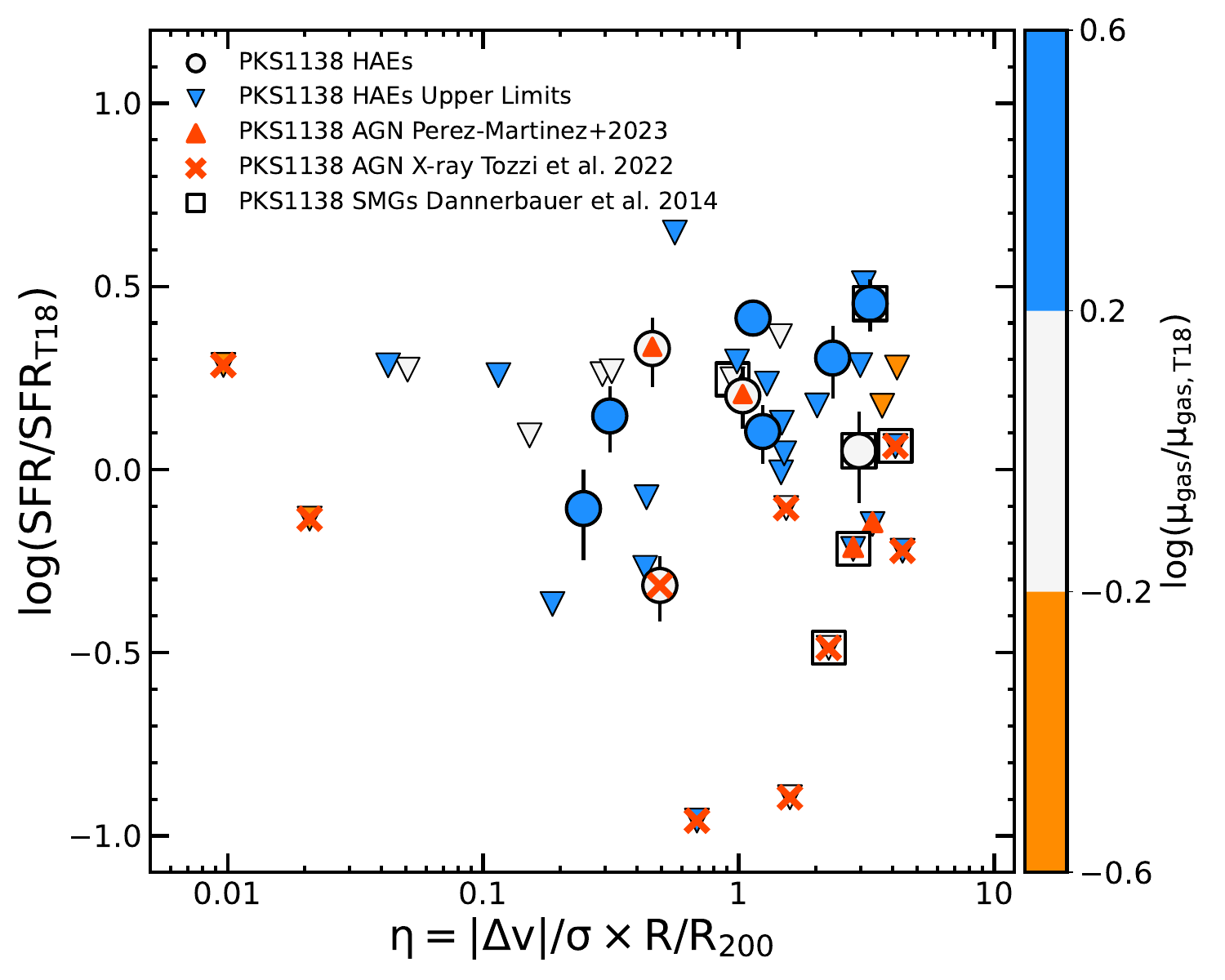}\par
\includegraphics[width=\linewidth]{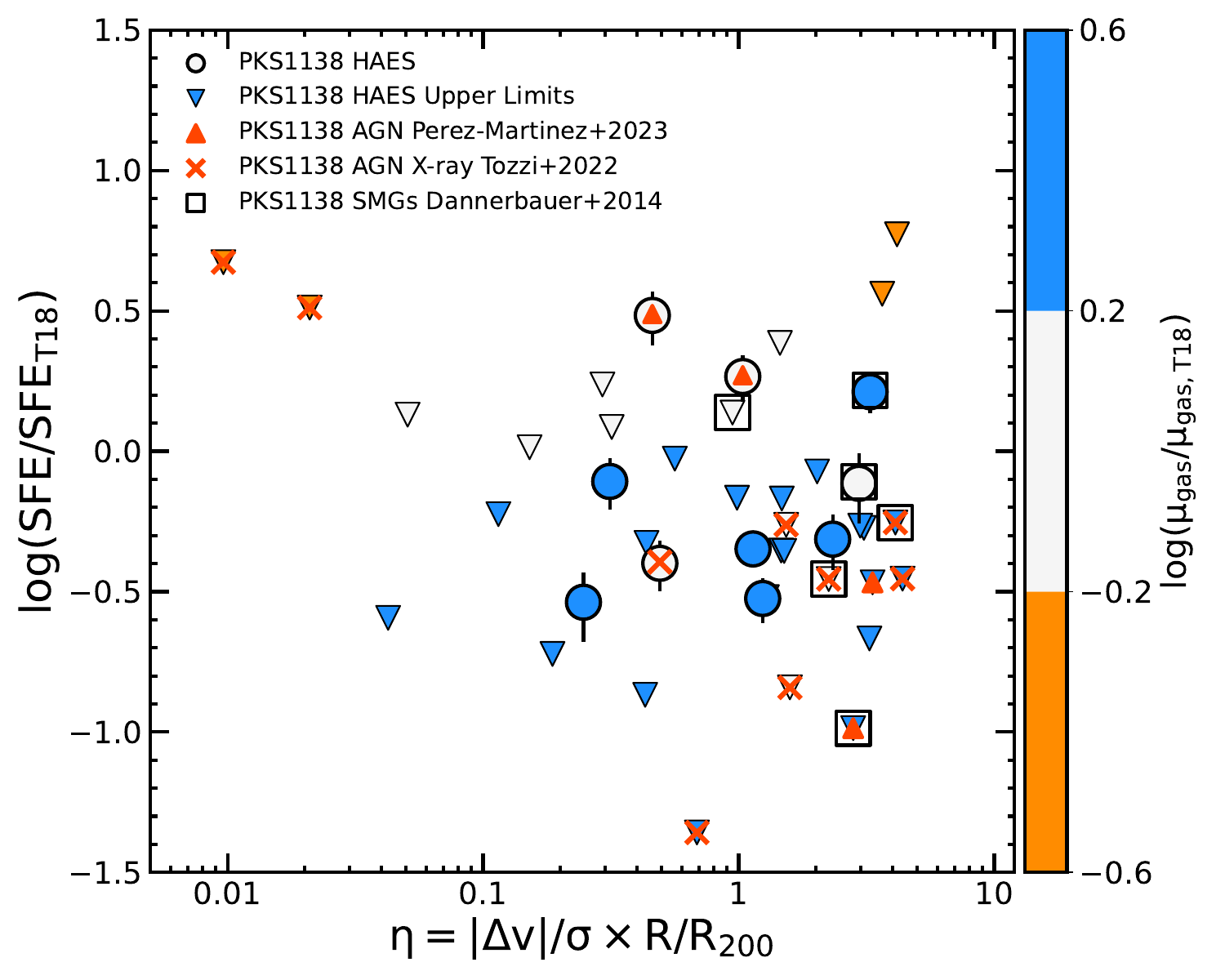}\par
\end{multicols}
    \caption{Left: Star formation rate offsets from the Main Sequence of \citet{Tacconi18} as a function of the environmental parameter $\mathrm{\eta}$. Right: Star formation efficiency offsets with respect to the scaling relation of \citet{Tacconi18} as a function of $\mathrm{\eta}$. The symbols and colors remain the same as in Fig.\,\ref{F:MolEnv}.}
        \label{F:MolEnv2}
\end{figure*}

In addition, Fig.\,\ref{F:MolEnv2} examines the relation between the star formation activities of our HAEs and their molecular gas reservoirs. The left panel of Fig.\,\ref{F:MolEnv2} shows a flat trend between the offsets from the main sequence of star formation and $\eta$. This means that regardless of the distance to the center of the cluster, most of the surveyed galaxies remain part of the Main Sequence even when reaching the cluster core, thus suggesting that the environment is not actively promoting or suppressing star formation over those objects. However, the color code shows again that galaxies in the outskirts of the protocluster have larger $\mathrm{\mu_{gas}}$ than their core counterparts, implying that the cold gas is gradually removed, heated up, or consumed as the galaxies approach the central regions. This can be seen more clearly in the right-hand panel of Fig.\,\ref{F:MolEnv2} where we display the offsets of the star formation efficiency (i.e., $\mathrm{SFE=SFR/M_{mol}=1/\tau_{dep}}$) with respect to the scaling relation of \cite{Tacconi18} as a function of $\eta$. Most objects display higher $\mathrm{\log(SFE/SFE_{T18})}$ values at lower $\mathrm{\eta}$ while this situation is inverted as we go further away from the protocluster core. However, our sample of HAEs is comprised of Main Sequence galaxies (Fig.\,\ref{F:MS} and the left panel in Fig.\,\ref{F:MolEnv2}) regardless of their position across the structure protocluster. This suggests that the observed increase in SFE towards the protocluster center is predominantly driven by the depletion of the gas reservoir in the central regions of the protocluster, in agreement with our previous results and those of \cite{Wang18}. Nevertheless, it is common to find more massive objects in overdensities compared to the coeval field as well as towards the center of a given overdensity compared to its outskirts. This stellar mass segregation (e.g., \citealt{Contini12}; \citealt{vandenBosch16}) is usually linked with the earlier formation times of these objects compared to those in less dense environments. Thus, stellar mass build-up may also play a role in the observed depletion of molecular gas trends towards the protocluster center on top of which environmental effects may be acting as discussed in Sect.\,\ref{S:Discussion}.

\begin{figure*}    
\centering
\begin{multicols}{3}
\includegraphics[width=\linewidth]{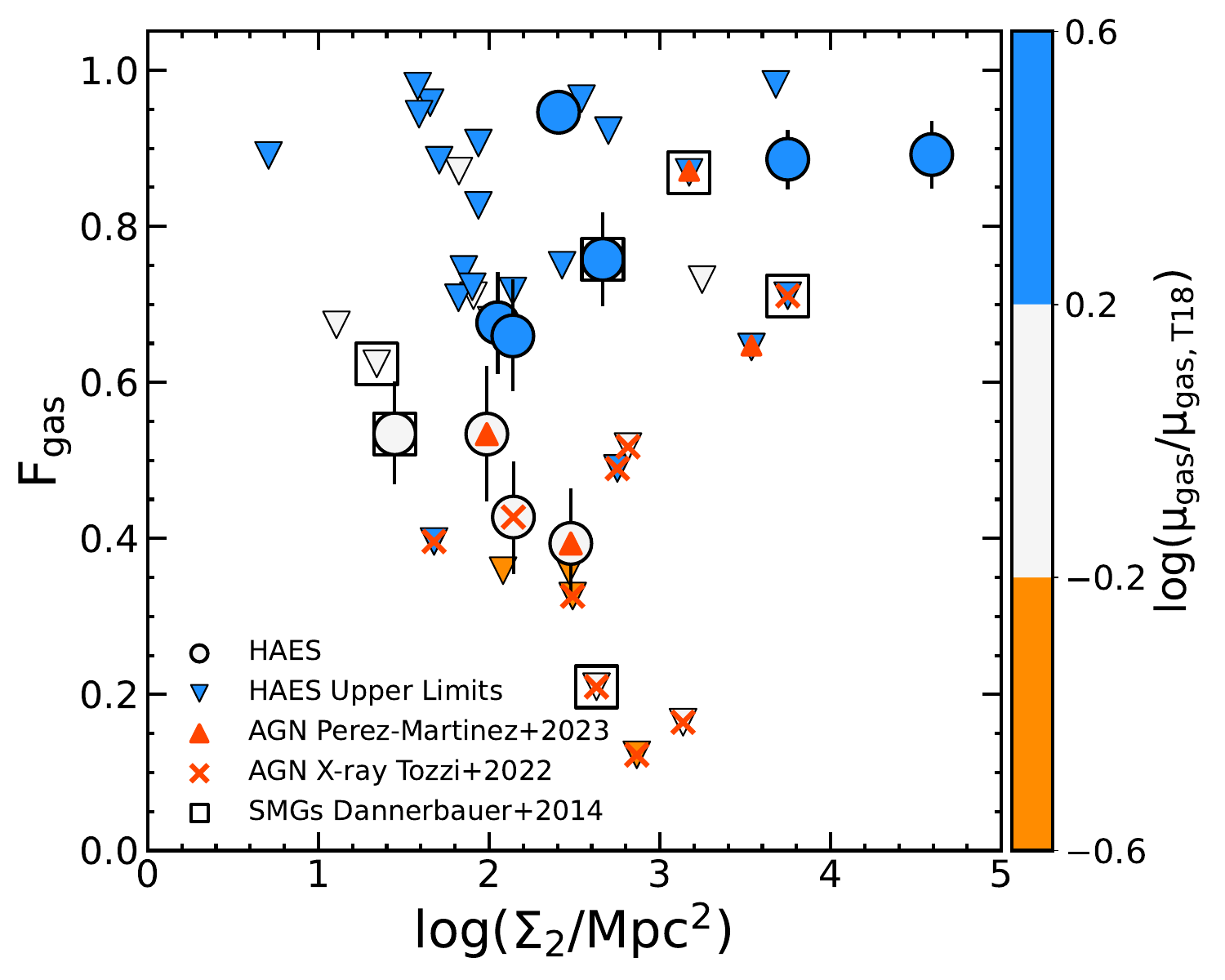}\par
\includegraphics[width=\linewidth]{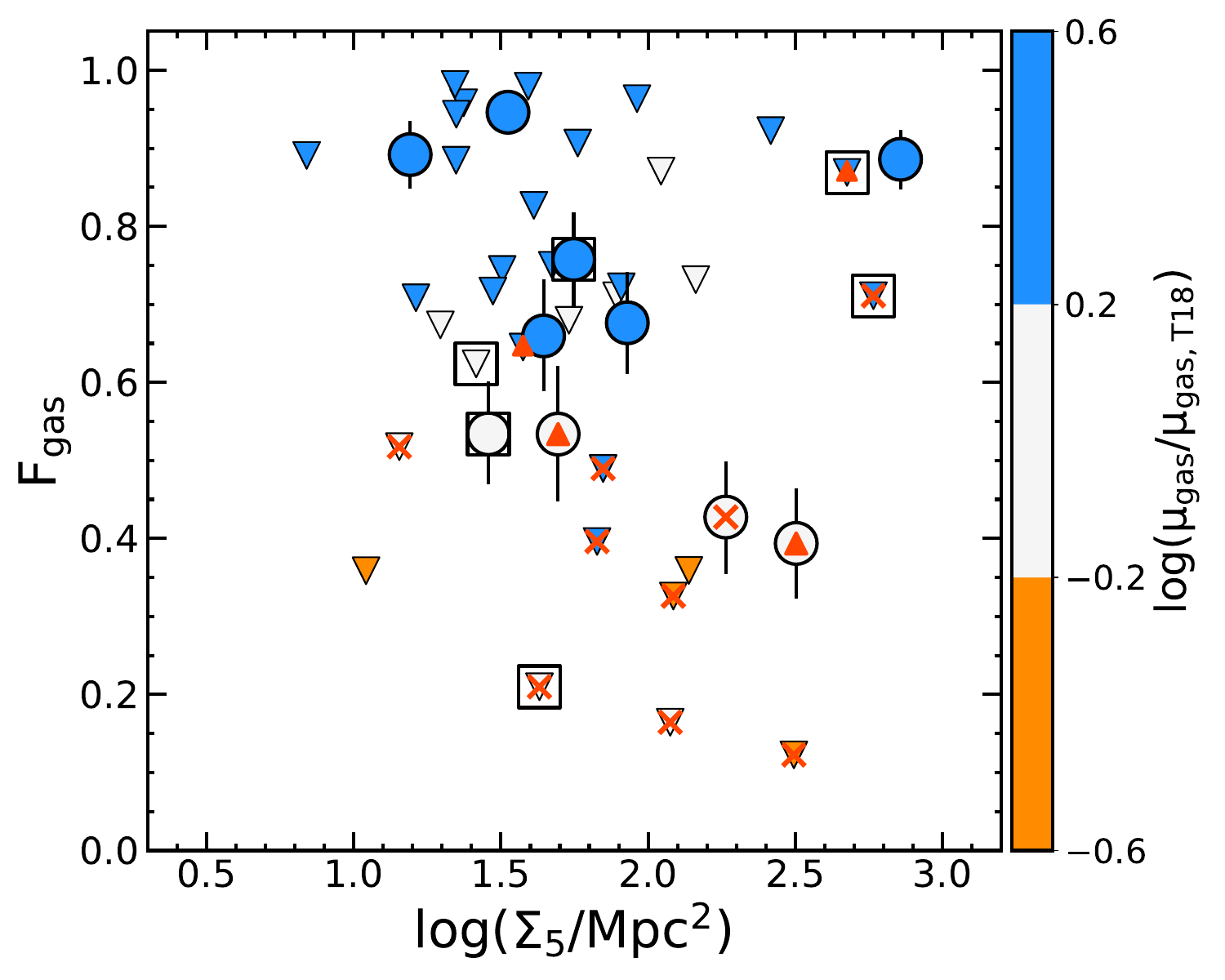}\par
\includegraphics[width=\linewidth]{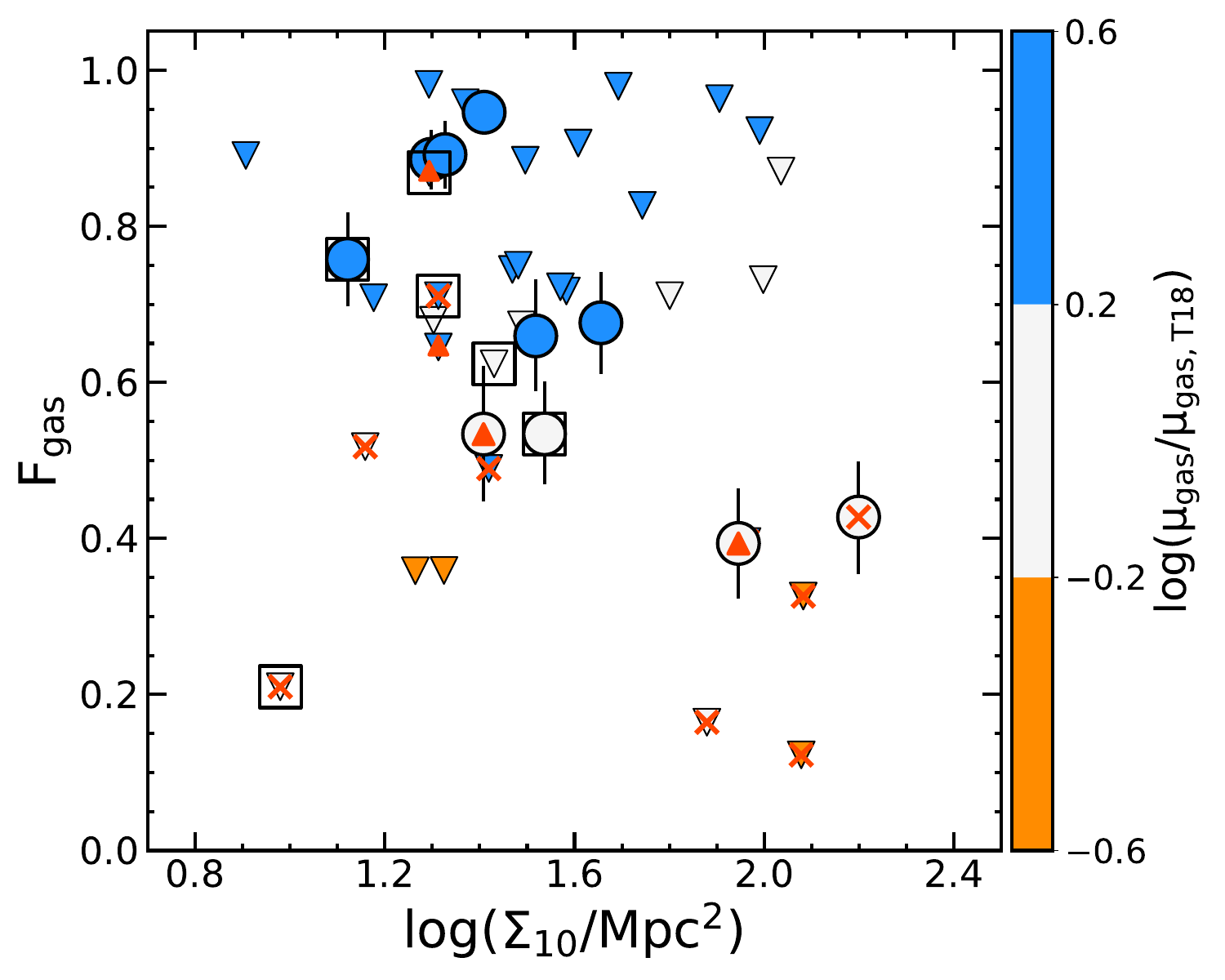}\par
\end{multicols}
    \caption{Gas fraction as a function of local density defined by the area enclosing two, five, and 10 neighboring galaxies ($\mathrm{\Sigma_2}$, $\mathrm{\Sigma_5}$, $\mathrm{\Sigma_{10}}$. The symbols and colors remain the same as in Fig.\,\ref{F:MolEnv}.}
        \label{F:MolEnv3}
\end{figure*}

Finally, Fig.\,\ref{F:MolEnv3} explore possible correlations between the gas fraction and the projected local surface density computed using the minimum area to enclose two, five, and ten galaxies (see Sect.\,\ref{SS:Environment}). This approach allows us to identify projected local density peaks (e.g., close companions or compact groups) across the structure of the protocluster. Our results show no clear correlation between $\mathrm{F_{gas}}$ and any of these three proxies, with galaxies displaying gas fractions above and below $\mathrm{F_{gas}=0.5}$ spreading across most of the local density range probed by these proxies. However, we acknowledge that the statistical limitations of our sample, the large number of upper limits, and the intrinsic bias of this proxy due to projection effects may contribute to erasing potential underlying correlations between these two quantities. 

\section{Discussion}
\label{S:Discussion}

This work has investigated the evolution of the molecular gas reservoir of a sample of main-sequence HAEs within the Spiderweb protocluster at $\mathrm{z=2.16}$. The sharp evolution of the total molecular gas fraction ($\mathrm{f_{gas}}$) at $\mathrm{\log M_*/M_\odot=10.5-11.0}$ shown in Fig.\,\ref{F:fgas} yields an evolutionary scenario composed of two main stages connected by a rapid transition at a given characteristic mass. In the low mass regime ($\mathrm{\log M_*/M_\odot\leq10.5}$), protocluster galaxies display molecular gas fractions close to unity despite having star formation activities already compatible with those of the main sequence (Fig.\,\ref{F:MS}). This implies that, while the bulk of their stellar mass remains to be formed in the future, the past and present stellar mass growth (i.e., SFR) has not depleted their gas reservoirs, suggesting that these objects keep replenishing their gas reservoirs through cold gas inflows from the cosmic web (\citealt{Tadaki19}). On the other hand, the high mass end of our HAEs displays $\mathrm{f_{gas}\leq0.4}$ with some objects going down to $\mathrm{f_{gas}\leq0.2}$ at $\mathrm{\log M_*/M_\odot=11.0-11.5}$. These objects represent the final stage of the in-situ stellar mass growth in galaxies, showing still active star formation albeit with clear signs of gas reservoir depletion, hinting at the formation of passive objects by the time this process is completed (e.g., \citealt{Falkendal19}; \citealt{Williams21}; \citealt{Whitaker21b}; \citealt{Zanella23}; \citealt{Blanquez-Sese23a}; \citealt{DEugenioC23}). The gap between these two phases is bridged by a population of rapidly transitioning galaxies at ($\mathrm{\log M_*/M_\odot=10.5-11.0}$) displaying $\mathrm{0.3<f_{gas}<0.9}$ and with a high fraction of AGN (12/20) at $\mathrm{\log M_*/M_\odot>10.5}$. Thus, the physical mechanism behind this transition must act on a relatively short timescale and predominantly act at and beyond the aforementioned stellar mass threshold, for which AGN feedback is a good candidate.

Furthermore, Fig.\,\ref{F:tdep} shows that most HAEs in our sample have depletion times ranging from 1 to 3 Gyrs, with a few of them showing even shorter time scales. This means that, in the absence of in-/outflows, the star-forming population that dominates the Spiderweb protocluster would become passive by $\mathrm{1<z<1.6}$. Observational evidence suggests that galaxy protoclusters start the surge of their red sequence by $\mathrm{z\approx2}$. In fact, past works in the Spiderweb protocluster already confirmed the presence of a handful of passively evolving galaxies near its core (\citealt{Kodama07}; \citealt{Zirm08}; \citealt{Tanaka10}; \citealt{Doherty10}; \citealt{Naufal24}) in line with other such studies in protoclusters (e.g., \citealt{Strazzullo16}; \citealt{Noordeh21}). However, most observational evidence suggests that the red sequence experiences significant growth in massive clusters at $\mathrm{1.5<z\lesssim2}$ (\citealt{Grutzbauch12}; \citealt{Fassbender14}; \citealt{Nantais16}; \citealt{Beifiori17}) until it becomes the dominant population within the cluster cores by $\mathrm{z=1}$ in agreement with our depletion times estimates. 

On the other hand, we have examined possible environmental impacts over our sample in Fig.\,\ref{F:MolEnv} and \,\ref{F:MolEnv2} through the molecular-to-stellar mass ratio ($\mathrm{\mu_{gas}}$), the star formation rate (SFR), and the star formation efficiency (SFE) as a function of the phase-space environmental parameter ($\mathrm{\eta}$, Eq.\,\ref{EQ:GlobalDensity}). Our findings suggest that the Spiderweb protoclusters host a large population of molecular gas-rich galaxies with lower star formation efficiency in its outskirts ($\mathrm{\eta>2}$) while this trend is reversed when approaching the protocluster core (i.e., lower $\mathrm{\mu_{gas}}$ and higher SFE at $\mathrm{\eta<0.4}$). However, our sample displays star formation activities in line with the main sequence regardless of their location across the cluster structure. This was previously reported by \cite{PerezMartinez23} and suggests that if environmental effects are at play, they do not have a significant impact on the star formation activities of the protocluster members for now and thus, any change in the molecular-to-stellar mass ratio $\mathrm{\mu_{gas}}$ is primarily driven by changes in the cold gas reservoir while star formation proceeds unaltered. We propose three different mechanisms to explain this process: changes in gas accretion, gas removal through environmental effects, and AGN activity.

\subsection{Changes between accretion regimes}
\label{SS:AccretionRegimes}

Environmental effects and gas accretion from the cosmic web (\citealt{Dekel06}; \citealt{Daddi22a}) could respectively hasten or delay the depletion of the gas reservoir. Current observational evidence can not rule out the presence of cold gas accretion within the Spiderweb protocluster. However, the recent discovery of a nascent ICM within its inner regions using Sunyaev-Zeldovich effect observations (\citealt{DiMascolo23}) suggests that if such accretion is still in place it may not proceed homogeneously across the protocluster structure. Thus, we would expect some galaxies to be fed through the so-called "cold streams in hot media" (\citealt{Dekel06,Dekel09a}) within the protocluster core. This process would progressively lose efficiency as the Spiderweb grows in mass and achieves virialization becoming a bona fide galaxy cluster. 

\begin{figure}
\includegraphics[width=\linewidth]{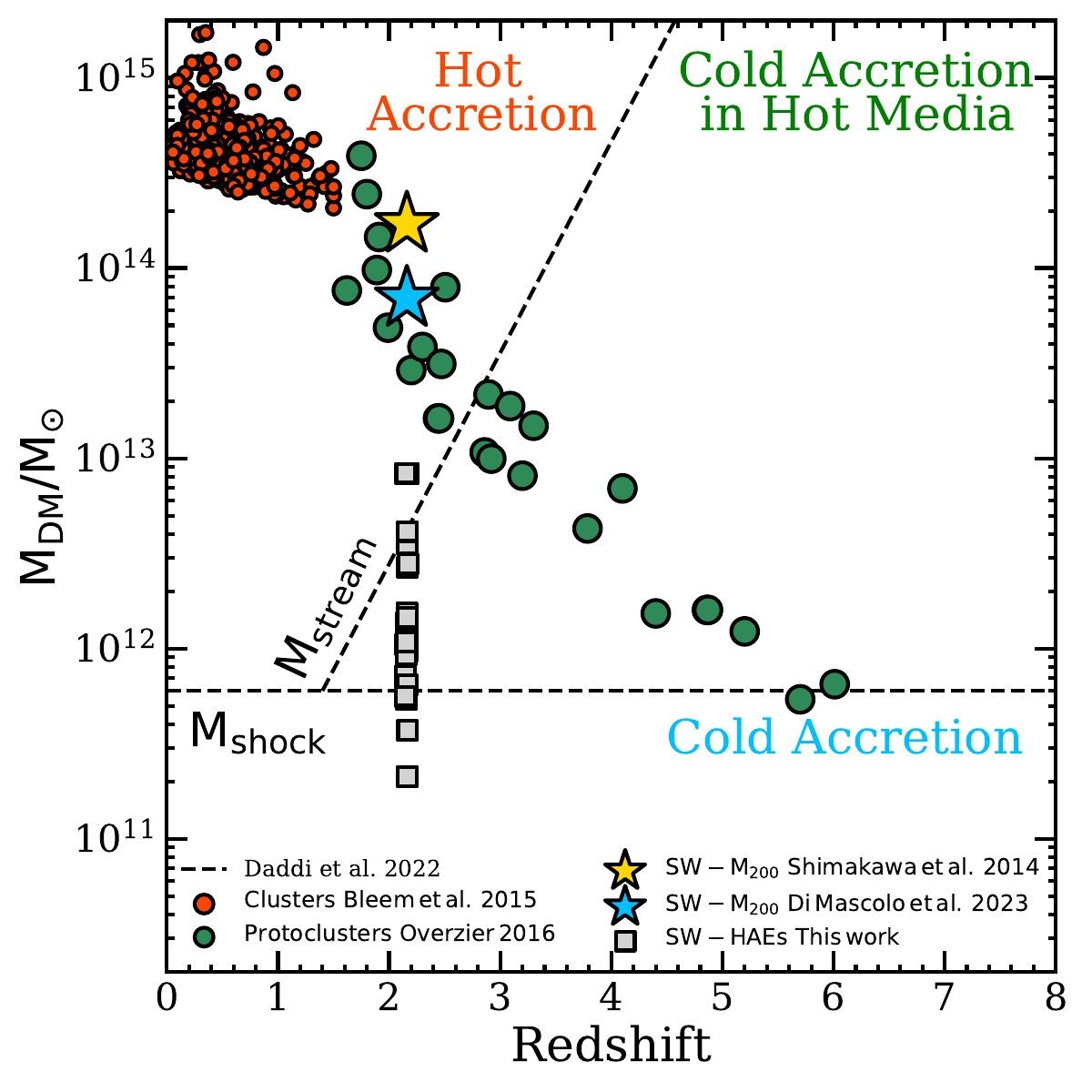}\par 
\caption{Accretion regime diagram. Colored stars display the dark matter halo mass of the Spiderweb protocluster derived using the SZ effect and dynamical estimates by \cite{DiMascolo23} and \cite{Shimakawa14} respectively. Grey squares depict the position of our sample of HAES after applying the stellar-to-halo mass relation in \cite{Behroozi19} for satellite galaxies (see Sect.\,\ref{SS:AccretionRegimes}). For comparison, we use green big circles to show a sample of protoclusters from \cite{Overzier16} while the small red circles represent a sample of SZ-detected $\mathrm{z<1.5}$ clusters from \cite{Bleem15}.}

\label{F:Accretion}
\end{figure}

Nonetheless, should cold accretion still be in place, we can roughly estimate its relevance in sustaining the star formation activities of the Spiderweb protocluster member galaxies through a series of simple assumptions. Several authors have provided practical approximations to estimate the expected baryonic accretion rate (BAR) within the $\mathrm{\Lambda CDM}$ cosmology for a given redshift and halo mass (e.g., \citealt{Genel08}; \citealt{Goerdt10}; \citealt{Dekel13}) based on hydrodynamical simulations. Recently, \cite{Daddi22a} investigated the transition between cold and hot accretion in a series of groups at the cosmic noon following Eq.\,5 in \cite{Goerdt10}, which we reproduce here:
\begin{equation}
\mathrm{BAR\approx137\times\left(\frac{M_{DM}}{10^{12}\,M_\odot}\right)^{1.15}\left(\frac{1+z}{4}\right)^{2.25}\,M_{\odot}yr^{-1}}
\label{EQ:BAR}
\end{equation}
Following the theory outlined by \cite{Dekel06}, dark matter haloes below a given mass threshold ($\mathrm{M_{DM}\lesssim M_{shock}=6\times10^{11}\,M_{\odot}}$) accrete all their baryons throughout cold streams. Thus Eq.\,\ref{EQ:BAR} approximately describes their total but also cold baryonic accretion rate. Above this limit, the cooling times are longer than dynamical times, and shocks can efficiently heat part of the gas. However, numerical simulations predict that cold accretion continues to penetrate at high-z ($\mathrm{z>z_{crit}}$) in the form of cold streams, surviving the shocks at the virial radii of massive halos up to $\mathrm{M_{DM}=M_{stream}}$ with $\mathrm{\log M_{stream}\approx\log M_{shock}+1.11\times(z-1.4)}$ and $\mathrm{z_{crit}=1.4}$ (\citealt{Dekel06}). However, recent observational works claim a flatter slope and lower $\mathrm{z_{crit}}$ for this boundary (e.g., \citealt{Daddi22b}). Dark matter haloes beyond this mass threshold ($\mathrm{M_{DM}\gtrsim M_{stream}}$) experience a smooth transition with decreasing cold accretion efficiency as their mass grows. \cite{Daddi22a} modeled such transition as $\mathrm{BAR_{cold}\approx BAR\times(M_{stream}/M_{DM})^{\alpha}}$ with $\mathrm{\alpha}$ being the modulation factor of such efficiency. Fig.\,\ref{F:Accretion} displays the Spiderweb protocluster accretion regime for the halo mass estimates using the Sunyaev-Zeldovich effect ($\mathrm{M_{500}=3.5\times10^{13}\,M_{\odot}yr^{-1}}$, \citealt{DiMascolo23}) and dynamical estimates ($\mathrm{M_{200}=1.7\times10^{14}\,M_{\odot}yr^{-1}}$, \citealt{Shimakawa14}). By comparison, we have also included the sample of SZ-detected clusters at $\mathrm{z<1.5}$ (\citealt{Bleem15}), a sample of protoclusters a $\mathrm{1.5<z<6.5}$ (\citealt{Overzier16}), and individual Spiderweb protocluster HAEs studied in this work after estimating their dark matter halo masses using the stellar-to-halo mass relation (SHMR) for satellite galaxies from \cite{Behroozi19}. At $\mathrm{z=2.16}$, this relation is essentially flat for $\mathrm{M_{*}>10^{10.8}\,M_{\odot}}$. We assign a maximum dark matter halo mass of $\mathrm{M_{DM}\approx10^{12.9}\,M_{\odot}}$ to the most massive HAEs in our sample, thus preventing unrealistically large variations of $\mathrm{M_{DM}}$ with small changes in stellar mass.  

Given these conditions and considering individual satellite dark matter haloes, a typical $\mathrm{M_*=10^{10}\,M_{\odot}}$ HAE will lie within a halo of $\mathrm{M_{DM}\approx10^{12}\,M_{\odot}}$. This value is well below $\mathrm{M_{stream}}$ at $\mathrm{z=2.16}$ and thus, its cold accretion is equal to a BAR of roughly $\mathrm{80\,M_{\odot}yr^{-1}}$. This level of cold gas accretion is comparatively higher than the expected gas consumption via star formation for a main sequence galaxy (\citealt{Speagle14}) of this stellar mass, with $\mathrm{SFR\approx30\,M_{\odot}yr^{-1}}$. Thus, the gas reservoir of galaxies within this stellar mass regime would be replenished as they grow in mass through star formation, keeping their gas fraction close to unity even if the efficiency of converting the cold gas into stars is only $\mathrm{50\%}$, as described by \cite{Dekel09a} models (see also \citealt{Daddi22a}). On the other hand, a typical $\mathrm{M_*=10^{11}\,M_{\odot}}$ galaxy will inhabit a dark matter halo of $\mathrm{M_{DM}\approx10^{13}\,M_{\odot}}$ just above $\mathrm{M_{stream}}$ at this redshift and thus lying within the so-called hot accretion regime (\citealt{Dekel06}). Here, the cold BAR depends on the modulation factor $\mathrm{\alpha}$ with recent studies suggesting that $\mathrm{\alpha\approx1.0\pm0.2}$ for haloes within approximately $\mathrm{\pm1.5\, dex}$ of $\mathrm{M_{stream}}$ (\citealt{Daddi22a,Daddi22b}), which would be the case of our most massive HAEs. Thus, their cold BAR is roughly $\mathrm{475^{+90}_{-77}\,M_{\odot}yr^{-1}}$ which would be enough to feed a $\mathrm{M_*=10^{11}\,M_{\odot}}$ main sequence star-forming galaxy displaying $\mathrm{SFR\approx180\,M_{\odot}yr^{-1}}$ when assuming the same cold gas-to-stellar mass conversion factor (i.e., $\mathrm{\sim50\%}$). This implies that in the absence of additional mechanisms such as gas removal or AGN feedback (see Sect.\,\ref{SS:RPS} and \ref{SS:AGNact}), the  gas that is consumed by star formation in these objects would be replenished by cold gas accretion at this redshift. Thus, gas deprivation (i.e., so-called starvation) due to dark matter halo growth of individual objects could not explain alone the low gas fractions reported in Sect.\,\ref{S:Results} for the massive end of our sample of HAEs. 

Nonetheless, the virial mass of the main dark matter halo associated with the Spiderweb protocluster is of the order of $\mathrm{\sim10^{14}\,M_{\odot}}$ with some of the satellite HAEs lying well within the virial radius ($\mathrm{R_{200}=0.53\,Mpc}$, \citealt{Shimakawa14}) and thus the accretion modulation factor, $\mathrm{\alpha}$, could be larger than unity thus diminishing the efficiency of cold accretion. To test this possibility, we inspect the gas fraction of the 12 HAEs (excluding the Spiderweb galaxy) lying within $\mathrm{R_{200}}$ in our sample ignoring projection effects. We find that seven of them are relatively low-mass galaxies and display $\mathrm{f_{gas}\gtrsim0.7}$ despite their proximity to the Spiderweb galaxy, which we take as the approximate center for the halo of this protocluster. The other five objects, on the other hand, have $\mathrm{M_*>10^{10.5}\,M_{\odot}}$ and $\mathrm{0.1<f_{gas}<0.6}$. However, we find another five objects in our sample with the same stellar mass and gas fraction ranges outside $\mathrm{R_{200}}$. These findings suggest that the changes between accretion regimes that satellite galaxies experience as they infall towards the core of the Spiderweb protocluster are unlikely to be the main mechanism behind the gas depletion we observe at the massive end of Fig.\,\ref{F:fgas}.

\subsection{Gas removal through environmental effects}
\label{SS:RPS}

Gas removal is a common effect discussed in the context of environmental effects and can obey different causes. Its most studied version in clusters at $\mathrm{z<1}$ is ram pressure stripping (RPS, \citealt{Gunn72}). Ram pressure stripping originates due to the dynamical friction between the hot dense gas of the ICM and the outer layers of the galaxies' ISM (\citealt{Boselli22}). The effectiveness of RPS is proportional to the square of the satellite's infalling velocity and to the density of the medium it goes through and thus, while RPS is very effective in quenching low-mass satellite galaxies in low-z clusters (e.g., \citealt{Lagana13}; \citealt{Roberts19}; \citealt{Mao22}), it is unlikely it has a significant impact in protoclusters due to the low-velocity proper motions and the relatively low density and likely heterogeneous distribution of a nascent ICM during the protocluster assembling phase. Nevertheless, recent works have measured an electron density of the order of $\mathrm{n_{e}\approx10^{-1}-10^{-2}\,cm^{-3}}$ and temperatures of $\mathrm{kT\approx0.6-2\,keV}$ for a thermal component within the inner $\mathrm{10-100\,kpc}$ of the Spiderweb protocluster (\citealt{Tozzi22b}; \citealt{Lepore24}) implying the presence of a cool-core for the first time at $\mathrm{z>2}$. Despite these values being similar to those found in low-z galaxy clusters (e.g., \citealt{Boehringer04,Boehringer07}; \citealt{Sanderson06,Sanderson09}), the small extent of the X-ray emission, the steepness of its electron density profile, and the asymmetries found in the X-ray and SZ emission suggest that we are witnessing the first stages of a nascent ICM, still far from virialization, as suggested by \cite{DiMascolo23} and \cite{Lepore24}. Thus, we would expect RPS to be a suitable mechanism to remove significant amounts of gas only within close proximity of the Spiderweb galaxy (e.g., \citealt{Kuiper11}) and in low mass galaxies (i.e., shallower gravitational potentials).

We explore this possibility by inspecting the 4 HAEs within our sample at a distance $\mathrm{<200\, kpc}$ from the Spiderweb galaxy. Two of them (IDs 1420 and 1498) are relatively low-mass galaxies ($\mathrm{\log M_*\leq10.1}$) displaying high gas fractions ($\mathrm{f_{gas}=0.7-0.9}$) arguing against the presence of RPS. In contrast, the other two objects (IDs 880 and 1501) are massive galaxies ($\mathrm{\log M_*\geq10.9}$) displaying lower gas fractions ($\mathrm{f_{gas}<0.3-0.5}$) and with confirmed AGN activity via X-ray emission, pointing towards different mechanism at play to explain the depletion of their gas reservoir. Furthermore, local universe studies in massive galaxy clusters have demonstrated that while loosely bound gas lying in the outskirts of low-mass satellite galaxies could be easily removed in large quantities via RPS (\citealt{Fumagalli14}; \citealt{Boselli16,Boselli22}) only extreme cases would have a similar effect over the denser and typically more centrally concentrated cold molecular gas (e.g., \citealt{Stevens21}; \citealt{Roberts23}). In that sense, the detection of extended ($\mathrm{\geq40\,kpc}$) molecular gas reservoirs in satellite galaxies across the protocluster structure (\citealt{Dannerbauer17}; \citealt{ChenZ24}) also argues against RPS being an efficient gas removal mechanism at this stage. We have checked the abundance of extended molecular gas reservoirs as described by \cite{ChenZ24} within our sample of HAEs finding that only 4 HAEs out of the 10 CO(1-0) emitters at $\mathrm{S/N>4}$ (i.e., $\mathrm{\sim40\%}$) within our sample fulfill this requisite either as robust (3) or tentative (1) extended molecular gas reservoirs. One of these robust detections (ID 902 in Table\,\ref{T:BigTable}) hosts a remarkable extended molecular gas reservoir previously studied by \cite{Dannerbauer17} and \cite{ChenZ24}. However, this ratio is similar but slightly inferior to the $\mathrm{\sim46\%}$ of extended molecular gas reservoir candidates (robust + tentative = 21) found by \cite{ChenZ24} out of the full COALAS sample of 46 CO(1-0) emitters. This suggests that H$\alpha$ emitters do not preferentially host large molecular gas reservoirs compared to the COALAS sample.

Alongside RPS, simulations have demonstrated that significant gas outflows can also be mediated by stellar feedback (e.g., \citealt{Feldmann15}; \citealt{Bahe15}; \citealt{Tollet19}). The efficiency of this mechanism on the removal of cold gas and thus star formation quenching depends on the mass of the host galaxy (i.e., the depth of the gravitational potential) and the magnitude of the ongoing star-formation episode (\citealt{Larson74}). Therefore, low-mass galaxies would be more likely to lose part of their gas reservoirs than their most massive counterparts (e.g., \citealt{El-Badry16}; \citealt{Christensen16}; \citealt{Keller16}; \citealt{Pontzen17}; \citealt{Bassini23}). The stronger gravitational pull of massive galaxies effectively confines the outflowing gas, causing it to fall back into the galaxy and form galactic fountains (\citealt{Shapiro76}). During the early Universe, however, starburst episodes were more frequent thus resulting in higher levels of supernova feedback which in extreme cases can eject large amounts of metal-enriched gas beyond the virial radius (\citealt{Kimmig24}), thus temporarily exhausting their star formation activities (\citealt{Remus24}). However, these same works show that stellar feedback alone is not enough to completely shut down star formation or prevent the inflow of new gas after a short period unless it couples with AGN activity and other environmentally driven processes (see also Sect.\,\ref{SS:AGNact}). The results presented in Sect.\,\ref{S:Results} do not allow us to discard that stellar feedback may be playing a role in individual cases, albeit the lack of a starbursting population (Fig.\,\ref{F:MS}) and the high gas fractions reported in the low-mass end of our sample (Fig.\,\ref{F:fgas}) argue against this mechanism being dominant on the evolution of the observed populations at this stage.

Finally, tidal interactions between infalling satellites and the global cluster potential, or frequent high-speed galaxy encounters (i.e., harassment, \citealt{Moore96,Moore98}) are often referred to as one of the mechanisms behind the morphological transformation of galaxies in overdense environments (\citealt{Moore99}; \citealt{Moss00}). In extreme cases, such processes may cause the partial removal of the gas reservoir of a galaxy through tidal stripping (e.g., \citealt{Spilker22}). The efficiency of harassment mainly depends on the geometry of the infalling satellite's orbit, the closest distance to the center of the potential, the cumulative number of fly-bys, and the intrinsic properties of the interacting galaxies (e.g., \citealt{Gnedin03}; \citealt{Mastropietro05}; \citealt{Smith10}; \citealt{Chang13}; \citealt{Bialas15}). Fig.\,\ref{F:fgas} and \ref{F:MolEnv} (left panel) show that the low gas fraction population in this work is dominated by massive galaxies ($\mathrm{\log M_*/M_\odot\geq10.5}$), spanning a large variety of projected clustercentric distances $\mathrm{0\lesssim R/R_{200}\lesssim3}$ with a high fraction of AGN candidates (12/20). While harassment is capable of inducing instabilities in the disk funneling gas towards the nucleus thus triggering AGN activity, the widespread distribution of the X-ray sources across the protocluster global and local density regimes (see also \citealt{PerezMartinez23}) argue against tidal stripping or harassment being an efficient mechanism to deplete the gas reservoir of the studied sample.

\subsection{Effects of AGN feedback}
\label{SS:AGNact}

Numerous observational and numerical studies have shown that gas depletion and subsequent quenching can also be a consequence of AGN feedback, expelling to the IGM part of the gas near the galaxy center that would otherwise be available for star formation while heating the remaining one (see \citealt{Heckman14} and references therein). Simulations predict that violent AGN outflows would be particularly efficient in removing gas from low-mass galaxies in very short timescales ($\mathrm{<100\,Myr}$), while this process would be inefficient in their massive counterparts, impacting only the galaxy's inner few kpc (\citealt{GuoH22}). Furthermore, \cite{Oppenheimer08} suggested that in this latter case, the large gravitational potential of the massive objects would be able to rebound (part of) the outflowing gas which will return to the galaxy after cooling down to be eventually recycled. Several authors have argued that a nascent ICM such as the one reported by \cite{DiMascolo23} in the Spiderweb protocluster would contribute to this scenario by impeding the outflowing gas leaving the galaxies, which would rain down over their disks contributing to enhancing their gas-phase metallicity (e.g., \citealt{Kulas13}; \citealt{Shimakawa18a}; \citealt{PerezMartinez23}). 
Gas depletion throughout violent outflows is thus unlikely to be the only (or the main) driver acting on the depletion of the molecular gas reservoir in our subsample of HAEs hosting AGN, as they largely trace the intermediate-to-high mass regime ($\mathrm{\log M_*/M_\odot>10.5}$) for $\mathrm{f_{gas}<0.7}$ (see Fig.\,\ref{F:fgas}).

Some hydrodynamical simulations suggest that the gradual depletion of cold gas is caused by the cumulative energy release from AGN feedback which, acting over sufficiently long timescales, end up heating the medium, suppressing cold accretion, and shutting down star formation (\citealt{Ma22}; \citealt{Ghodsi24}). In this scenario, black hole mass ($\mathrm{M_{BH}}$) would be an accurate predictor for gas depletion due to AGN feedback as it traces the black hole's growth history (\citealt{Piotrowska22}) while current activity indicators such as the Eddington ratio would play a secondary role except for extreme growth episodes (e.g., \citealt{GuoH22}). Interestingly, \cite{Zinger20} found that a black hole mass scale of $\mathrm{M_{BH}=10^{7.5}-10^{8.5}}$ is linked to drastic changes in several key physical parameters (e.g., cooling times, gas fraction and star formation) driving the transformation between low-mass, gas-rich, star-forming galaxies to massive gas-depleted quiescent using the IllustrisTNG project suite of simulations (\citealt{Springel10}; \citealt{Pillepich18}). A similar $\mathrm{M_{BH}}$ scale was found by \cite{Ma22} and \cite{Ghodsi24} when examining the depletion of the molecular gas reservoir and quenching of star formation for both IllustrisTNG and SIMBA (\citealt{Dave16,Dave19}) or by \cite{Bower18} using the EAGLE simulation (\citealt{Schaye15}). This $\mathrm{M_{BH}}$ roughly corresponds with a stellar mass of $\mathrm{M_{*}\approx10^{10.5}\,M_{\odot}}$ assuming an almost invariant stellar to black hole mass relation at $\mathrm{0<z<2.0}$ (\citealt{Reines15}; \citealt{Suh20}; \citealt{Tanaka24}).  This stellar mass threshold and sharp transition of properties resembles our Fig.\,\ref{F:fgas} results where the molecular gas fraction estimates of our HAEs drops from $\mathrm{f_{gas}\approx0.9}$ at $\mathrm{M_{*}\lesssim10^{10.5}\,M_{\odot}}$ to $\mathrm{f_{gas}\approx0.3}$ at the massive end. Furthermore, 12 out of the 20 objects at $\mathrm{M_{*}\gtrsim10^{10.5}\,M_{\odot}}$ show signs of current AGN activity throughout X-ray emission or emission line diagnostics, reinforcing the idea that the abrupt change of gas properties may be linked to AGN activity. This possibility is also discussed by \cite{Kolwa23} for a sample of field massive ($\mathrm{M_{*}\gtrsim10^{11}\,M_{\odot}}$) and radio-loud AGN host at $\mathrm{z=2.9-4.3}$. These authors find high levels of gas depletion ($\mathrm{f_{gas}<0.2}$) through the [C\,{\sc{i}}](1-0) molecular gas tracer in contrast with star-forming or SMG coeval field samples. Nevertheless, this comparison between observational and simulated results might be limited by several factors. For example, the various specific implementations of AGN feedback between theoretical works (\citealt{Hopkins15}; \citealt{Weinberger17}), which are tuned to be the main quenching mechanism of massive objects in SIMBA and the TNG simulations. In addition, several of the aforementioned works focus on the evolution of central galaxies at $\mathrm{z=0}$, thus minimizing the impact of additional environmental effects and neglecting intrinsic differences between the star-forming population of the local Universe and the Cosmic Noon.

On the other hand, \cite{Kimmig24} proposes a somewhat different scenario using the Magneticum Pathfinder simulation suite. In this case, the best predictor for the quenching of massive high redshift galaxies would depend on the timescale scrutinized. On short timescales ($\mathrm{<\,500\,Myrs}$) galaxies may be quenched by the combined feedback of a violent starburst followed by a powerful episode of AGN activity. These events would be mediated either by strong accretion or gravitational interactions, able to trigger star formation and funnel gas to the central engine simultaneously. In this case, the best predictor for quenching is the black hole to stellar mass ratio while the correlation with the environment is subdominant. Thus, the population of recently quenched or soon-to-be-quenched massive satellite galaxies identified in $\mathrm{z>2}$ protoclusters during the last years (\citealt{Kubo21,Kubo22}; \citealt{McConachie22}; \citealt{Ito23}; \citealt{Jin24}; \citealt{ShiD24}) may have predefined their main quenching mechanism independently of the environment they reside in at the time of their discovery. However, on quenching timescales longer than 1 Gyr the environment becomes the best predictor for quenching as the suppression of cold gas accretion eventually shuts down star formation (overconsumption, \citealt{McGee14}) and prevents future episodes of rejuvenation as discussed by \cite{Remus24} using the same suit of simulations. Interestingly, this environmentally driven lack of accretion has different origins for massive central and satellite galaxies. Massive central quenched galaxies end up residing in underdense regions and are unable to accrete enough fuel to overcome the heating of their haloes and rekindle. However, the gas accretion of massive satellite galaxies is stopped due to the growth of the nascent ICM within the overdensity they reside in, and due to the onset of classic environmental effects such as RPS.

Our results in the Spiderweb protocluster do not show a synchronous decrease in SFR as the molecular gas reservoir is depleted (Figures\,\ref{F:MS} and \ref{F:fgas}) in contrast with the expectations from \cite{Zinger20} and \cite{Ma22}. At the same time, we do not detect ongoing starburst activity in our massive galaxies based on the available $\mathrm{H\alpha}$-based SFR, albeit we do see signs of AGN activity for a significant number of them. One possibility is that the ISM conditions for star formation may remain relatively stable while the gas reservoir is exhausted over time. In this scenario, quenching would be the consequence of overconsumption and thus felt by the galaxy only after the gas fraction reaches a critically low value. During this process, gas heating through previous and ongoing stellar and AGN feedback would prevent the incoming gas from cooling (\citealt{Li15,Li17}; \citealt{Dubois13}), thus ensuring the gradual depletion of the molecular gas reservoir. Taking as a reference that the transition phase in $\mathrm{f_{gas}}$ happens for galaxies at $\mathrm{10.5<\log M_*/M_\odot<11}$ and that the typical SFR for a main sequence galaxy at $\mathrm{\log M_*/M_\odot=10.5}$ is $\mathrm{SFR\approx30-100\,M_\odot/yr}$ we obtain that the time required to bridge this mass growth by in-situ star formation oscillates between 0.7 and 2.3\,Gyrs, which is shorter but not far from the typical depletion times measured in our sample (Fig.\,\ref{F:tdep}, $\mathrm{\tau_{dep}=1-3\,Gyr}$). Naturally, further environmental processes and their interrelations may extend or shorten this timescale. For example, as the Spiderweb protocluster continues growing its nascent ICM (\citealt{DiMascolo23}), RPS may become a more relevant and widespread phenomenon, which coupled with AGN and stellar feedback episodes could rapidly remove the gas reservoir of some satellite galaxies (e.g., \citealt{Bahe15}; \citealt{Noble19}). However, its efficiency would depend on the stellar mass of the galaxy, geometrical factors, and the local physical conditions on its orbit through the ICM (see \citealt{Boselli22} for a review). At the same time, mergers could both accelerate the mass growth by bringing external material or inhibit it by directly expelling gas or triggering AGN activity depending on the specific gas and stellar mass properties of the satellite galaxies involved (\citealt{Pontzen17}; \citealt{Sparre17}; \citealt{Lagos18a}).

Nevertheless, our SFRs rely on narrow-band and spectroscopic $\mathrm{H\alpha}$ flux measurements carried out by \cite{Shimakawa18a} and \cite{PerezMartinez23} respectively. This tracer is unable to capture heavily obscured star formation episodes common in SMGs. However, only 5 objects within our sample are identified as such (\citealt{Dannerbauer14}) and they reside in the protocluster outskirts. In addition, these previous works assumed a star-forming origin as the main ionization source for $\mathrm{H\alpha}$. This assumption is somehow reinforced by the low fraction of broad-line emitters identified by \cite{PerezMartinez23}, with only 2 broad-line AGN out of 39 spectroscopically analyzed HAEs. However, the X-ray identifications reported in \cite{Tozzi22a} and the multi-wavelength SED analysis carried out by \cite{Shimakawa24} over these same sources suggest that a fraction of the $\mathrm{H\alpha}$ flux of these sources may be contaminated by the ionization from the central engine which could result in the overestimation of our SFR for this sources, thus placing them up to $\mathrm{\sim1\,dex}$ below the main sequence of star formation and in qualitative agreement with the predictions from the numerical simulations (\citealt{Zinger20}).

Regardless of AGN feedback being (or not) the dominant factor for the cold gas depletion and quenching of star formation at the massive end of our sample, it remains unclear if the cumulative growth of the supermassive black hole or the triggering event for its current activity are linked to environmental factors such as gas angular momentum loss caused by enhanced galaxy-galaxy interactions in protoclusters (e.g., \citealt{Lotz13}; \citealt{Hine16}; \citealt{Watson19}; \citealt{LiuS23}; \citealt{Naufal23}; \citealt{Shibuya24}) or it rather merely responds to secular processes (i.e., enhanced accretion) acting over the assembly history of massive galaxies which are statistically amplified by the fact that protoclusters tend to host larger number counts of massive objects than the coeval field (e.g., \citealt{Nantais16}; \citealt{Shimakawa18b}; \citealt{Hill22}). During the last years, several teams hinted at diverse quenching mechanisms for passively evolving galaxies residing within protoclusters albeit current observational evidence does not allow to pinpoint a purely environmental origin (e.g., \citealt{Kalita21}; \citealt{Kubo21}; \citealt{McConachie22}; \citealt{Ito23}; \citealt{ShiD24}; \citealt{Jin24}). In the Spiderweb protocluster, AGNs are widely distributed in terms of phase-space (Fig.\,\ref{F:MolEnv}) as well as projected local densities (Fig.\,\ref{F:MolEnv3}, see also \citealt{PerezMartinez23}) making it difficult to establish a causal link between their local environment and their ongoing AGN activity. Furthermore, current broad-band photometry in the NIR lacks the resolution to identify tidal tails or heavily disturbed morphologies that could act as the smoking gun for recent gravitational interactions (\citealt{Jin24}). 

\section{Conclusions}
\label{S:Conclusions}

In this work, we have analyzed the molecular gas properties of a sample of HAEs within the Spiderweb protocluster using CO(1-0) measurements and upper limits from the COALAS survey as a proxy to weigh their cold gas reservoir. We have examined $\mathrm{f_{gas},\,\tau_{dep}}$, SFR, and SFE in our sample and put in compare our results with coeval scaling relations and observational works both in the field and protoclusters (Sect.\,\ref{S:Results}). Finally, we have discussed the implications of these results on different scenarios of galaxy evolution during the early stages of protocluster assembly (Sect.\,\ref{S:Discussion}). In the following paragraphs, we summarize the main conclusions of this work:

\begin{enumerate}
\item We have identified 43 spectroscopically confirmed HAEs within the COALAS ATCA CO(1-0) footprint and obtained CO(1-0) detection for 10 objects in agreement with \cite{Jin21}. In addition, we have obtained 26 CO(1-0) upper flux limits for the rest of our sample. 
\item Our protocluster sample displays a very sharp transition from high to low molecular gas fractions ($\mathrm{f_{gas}}$) values at stellar masses $\mathrm{\log M_*/M_\odot=10.5-11.0}$, in agreement with previous coeval observational CO(1-0) works in protoclusters (\citealt{Wang18}) and the field (\cite{Riechers20a}. The distribution of these sources in the $\mathrm{f_{gas}-M_{*}}$ plane can be fitted by a logistic function (Eq.\,\ref{EQ:logistic}) as suggested by \cite{Popping12}. This is in contrast with the smoother functions reported by \cite{Sargent14} and \cite{Tacconi18}.
\item More than half (12/20) of the HAEs with $\mathrm{\log M_*/M_\odot\geq10.5}$ host X-ray (\citealt{Tozzi22a}) or emission line diagnostics indicating the presence of an AGN (\citealt{PerezMartinez23}). Furthermore, the stacking of these sources reveals a lower gas fraction value ($\mathrm{F_{gas}=0.38\pm0.05}$) than that of equally massive sources but without signs of AGN activity ($\mathrm{F_{gas}=0.55\pm0.04}$). This hints that nuclear activity significantly contributes to molecular gas depletion beyond this mass limit. 
\item The measured depletion times ($\mathrm{\tau_{dep}}$) show that, in the absence of inflows, most of these HAEs will consume their gas reservoir via star formation in $\mathrm{1-3\,Gyrs}$ (i.e., by $\mathrm{1<z<1.6}$), concurrently with the observational evidence suggesting that massive galaxy clusters establish their red sequence at $\mathrm{1.5<z\lesssim2}$ which will end up dominating their cores by $\mathrm{z=1}$. Environmental effects during this period could further shorten the depletion time while cold gas accretion could delay it. 
\item Our environmental analysis suggests that the fraction of HAEs with higher molecular-to-stellar mass ratios ($\mathrm{\mu_{gas}}$) increases towards the outskirts of the cluster (Fig.\,\ref{F:MolEnv}). The opposite trend is found in terms of SFE, similarly to \citet{Wang18}. However, SFR shows no clear correlation as a function of phase space ($\mathrm{\eta}$) in this protocluster (Fig.\,\ref{F:MolEnv2}, see also \citealt{PerezMartinez23}). This implies that the changes in SFE are predominately driven by decreasing molecular gas reservoirs while star formation activity proceeds unaffected for most galaxies regardless of their location within the protocluster structure.
\item We propose three possible mechanisms to explain the depletion of the molecular gas reservoir: changes in gas accretion due to the halo growth, gas removal through environmental effects, and AGN activity. The sharp change in gas fraction at $\mathrm{\log M_*/M_\odot=10.5-11.0}$ suggests that the dominant mechanism predominately acts in the high stellar mass regime.
 Thus, we propose AGN gas heating coupled with overconsumption as the likely mechanism behind this behavior. As the nascent ICM grows, the efficiency of gas accretion diminishes. On top of this, AGN heating contributes to the confinement of the host galaxy while allowing its star formation activities to continue in a relatively unaltered way. As a result, the galaxy will end up depleting its gas reservoir and quenching. This scenario is supported by the high AGN fraction between massive galaxies in this protocluster. On the other hand gas removal would have a secondary role in this stellar mass range, as it is typically more efficient on lower-mass galaxies (e.g., ram pressure or tidal stripping) at later cosmic epochs when clusters have fully virialized and developed their ICM.
 
\end{enumerate}
 
Overall, this work has conducted one of the most comprehensive studies of star formation and gas consumption in a protocluster during the cosmic noon despite the spatial resolution constraints of both ATCA submillimetre and NIR ground-based observations. The recent advent of ALMA Band-1 in Cycle 10 provides new opportunities to examine and replicate our findings using CO(1-0) at higher spatial resolution and across protoclusters at different evolutionary stages during this epoch. Furthermore, the integration of ALMA with the exceptional NIR and MIR capabilities of the James Webb Space Telescope (JWST) will enable a deeper understanding of some of the remaining challenges for galaxy evolution in early overdense environments in the near future. These include determining when, where, and how intensely star-forming activities occur within protocluster members, and what roles the environment plays in enhancing or supressing them during the early stages of cluster assembly.

\begin{acknowledgements}
The Australia Telescope Compact Array is part of the Australia Telescope National Facility (grid.421683.a) which is funded by the Australian Government for operation as a National Facility managed by CSIRO. We acknowledge the Gomeroi people as the traditional owners of the Observatory site. This research is based in part on observations collected at the European Organisation for Astronomical Research in the Southern Hemisphere under ESO programme 095.A-0500(B). This research is based in part on data collected at the Subaru Telescope, which is operated by the National Astronomical Observatory of Japan (NAOJ). We are honored and grateful for the opportunity of observing the Universe from Maunakea, which has cultural, historical, and natural significance in Hawaii. This work is based in part on observations made with the Spitzer Space Telescope, which was operated by the Jet Propulsion Laboratory, California Institute of Technology under a contract with NASA. This research is based in part on observations made with the NASA/ESA Hubble Space Telescope obtained from the Space Telescope Science Institute, which is operated by the Association of Universities for Research in Astronomy, Inc., under NASA contract NAS 5–26555. The National Radio Observatory is a facility of the National Science Foundation operated under cooperative agreement by Associated Universities, Inc. This research made use of Astropy,\footnote{http://www.astropy.org} a community-developed core Python package for Astronomy \citep{Astropy13, Astropy18}. JMPM acknowledges funding from the European Union’s Horizon-Europe research and innovation programme under the Marie Skłodowska-Curie grant agreement No 101106626. HD, ZC, YZ, and JMPM acknowledge financial support from the Agencia Estatal de Investigación del Ministerio de Ciencia e Innovación (AEI-MCINN) under grant (La evolución de los c\'umulos de galaxias desde el amanecer hasta el mediod\'ia c\'osmico) with reference (PID2019-105776GB-I00/DOI:10.13039/501100011033). HD, YZ, and JMPM acknowledge financial support from the Ministerio de Ciencia, Innovaci\'on y Universidades (MCIU/AEI) under grant (Construcci\'on de c\' umulos de galaxias en formaci\'on a trav\'es de la formaci\'on estelar oscurecida por el polvo) and the European Regional Development Fund (ERDF) with reference (PID2022-143243NB-I00/DOI:10.13039/501100011033). C.D.E. acknowledges funding from the MCIN/AEI (Spain) and the "NextGenerationEU"/PRTR (European Union) through the Juan de la Cierva-Formación program (FJC2021-047307-I). YZ acknowledges the support from the China Scholarship Council (202206340048). M.S.P. acknowledges the support of the Spanish Ministry of Science, Innovation and Universities through the project PID-2021-122544NB-C43. 
\end{acknowledgements}

\bibliographystyle{aa}
\bibliography{aa.bib}

\begin{appendix} 

\section{Summary Table}

\renewcommand{\arraystretch}{1.5}
\begin{sidewaystable*}
\caption{Physical properties of the Spiderweb protocluster HAES for which we have obtained ATCA CO(1-0) measurements. Coordinates (RA and DEC) are shown in degrees. The redshifts (z) and oxygen abundances (12+log(O/H)) are based on the $\mathrm{H\alpha}$ and [N{\sc{ii}}] emission lines (\citealt{Shimakawa15}; \citealt{PerezMartinez23}). Stellar masses ($\mathrm{M_*}$), star formation rates (SFR), molecular gas masses ($\mathrm{M_{mol}}$), molecular gas fractions ($\mathrm{f_{gas}}$), and depletion times ($\mathrm{\tau_{dep}}$) have been computed following the steps outlined in Sect.\,\ref{S:Methods}. The last column highlights the presence of X-ray emission within 1\arcsec of the positions reported by \protect\cite{Tozzi22a} using Chandra 0.5-10 keV observations in this field.}
\begin{tabular}{cccccccccccccccccc}
\hline
\noalign{\vskip 0.0cm}
ID & RA  & DEC   & z  &  $\mathrm{\log\,M_*/M_\odot}$  & SFR & $\mathrm{12+\log(O/H)}$ & $\mathrm{\log\,M_{mol}/M_\odot}$ & $\mathrm{f_{gas}}$ & $\mathrm{\tau_{dep}}$ & X-ray \\
   & (J2000) & (J2000) & & & ($\mathrm{M_\odot/yr}$) & & & & ($\mathrm{Gyr}$) & \\
\noalign{\vskip 0.1cm}
\hline
210 & 175.1555833 & -26.5048056 & 2.1684 & $11.16^{+0.09}_{-0.11}$ & $101\pm8$ & $8.65\pm0.06$ & $<11.14$ & $<0.49$ & $<1.4$ & Yes \\ 
229 & 175.1573750 & -26.4867500 & 2.1574 & $10.98^{+0.11}_{-0.14}$ & $66\pm18$ & - & $<11.37$ & $<0.71$ & $<3.5$ & -  \\ 
255 & 175.1590000 & -26.5067500 & 2.1437 & $10.50^{+0.08}_{-0.10}$ & $101\pm22$ & - & $<10.25$ & $<0.36$ & $<0.18$ & - \\
298 & 175.1642500 & -26.5068333 & 2.1626 & $10.42^{+0.14}_{-0.20}$ & $70\pm13$ & $8.52\pm0.12$ & $<10.88$ & $<0.74$ & $<1.1$ & - \\
323 & 175.1655417 & -26.4792222 & 2.1620 & $10.98^{+0.08}_{-0.09}$ & $96\pm29$ & - & $<11.01$ & $<0.52$ & $<1.06$ & Yes \\
343 & 175.1670833 & -26.4963889 & 2.1609 & $10.11^{+0.12}_{-0.17}$ & $35\pm6$ & $8.66\pm0.07$ & $11.00\pm0.04$ & $0.89^{+0.05}_{-0.05}$ & $2.9\pm1.1$ & -  \\ 
408 & 175.1736667 & -26.4817222 & 2.1634 & $10.37^{+0.11}_{-0.15}$ & $99\pm22$ & - & $<10.68$ & $<0.67$ & $<0.49$ & - \\
457 & 175.1785833 & -26.4686944 & 2.1622 & $9.79^{+0.10}_{-0.13}$ & $27\pm4$ & $8.44\pm0.08$ & $<11.01$ & $<0.94$ & $<3.8$ & -  \\ 
478 & 175.1809167 & -26.4936944 & 2.1463 & $9.45^{+0.10}_{-0.13}$ & $10\pm3$ & - & $<10.68$ & $<0.88$ & $<3.7$ & - \\
503 & 175.1839167 & -26.4788889 & 2.1634 & $9.72^{+0.13}_{-0.18}$ & $28\pm5$ & - & $<10.71$ & $<0.91$ & $<1.4$ & - \\
511 & 175.1843750 & -26.4853056 & 2.1583 & $11.29^{+0.09}_{-0.11}$ & $27\pm13$ & - & $<10.58$ & $<0.16$ & $<1.4$ & Yes \\ 
527 & 175.1853333 & -26.4890556 & 2.1620 & $10.99^{+0.07}_{-0.08}$ & $14\pm5$ & - & $<10.81$ & $<0.40$ & $<3.0$ & Yes \\
586 & 175.2637083 & -26.4807778 & 2.1492 & $10.17^{+0.11}_{-0.15}$ & $58\pm19$ & $8.69\pm0.18$ & $<11.08$ & $<0.89$ & $<2.1$ & - \\ 
647$^b$ & 175.2599167 & -26.4625278 & 2.1510 & $11.59^{+0.10}_{-0.13}$ & $114\pm29$ & $8.80\pm0.07$ & $<11.01$ & $<0.21$ & $<0.9$ & Yes \\ 
779 & 175.2505833 & -26.4823056 & 2.1665 & $10.66^{+0.08}_{-0.09}$ & $106\pm14$ & $8.48\pm0.08$ & $<10.41$ & $<0.36$ & $<0.2$ & -  \\ 
790 & 175.2484167 & -26.5108611 & 2.1645 & $10.74^{+0.10}_{-0.13}$ & $231\pm38$ & $8.58\pm0.08$ & $11.06\pm0.02$ & $0.68^{+0.09}_{-0.09}$ & $0.5\pm0.1$ & - \\ 
876 & 175.1952083 & -26.4780833 & 2.1636 & $8.88^{+0.15}_{-0.23}$ & $15\pm2$ & - & $<10.56$ & $<0.98$ & $<2.5$ & -  \\ 
880 & 175.1944583 & -26.4862222 & 2.1663 & $10.90^{+0.07}_{-0.09}$ & $52\pm11$ & - & $10.88\pm0.03$ & $0.49^{+0.06}_{-0.06}$ & $1.5\pm0.6$ & Yes \\ 
902$^a$ & 175.1919167 & -26.4864722 & 2.1490 & $11.37^{+0.10}_{-0.13}$ & $511\pm110$ & $8.78\pm0.06$ & $11.13\pm0.02$ & $0.37^{+0.07}_{-0.07}$ & $0.3\pm0.1$ & - \\
903 & 175.1921667 & -26.4902222 & 2.1553 & $10.35^{+0.11}_{-0.14}$ & $77\pm7$ & $8.36\pm0.06$ & $<10.79$ & $<0.73$ & $<0.8$ & - \\
911$^b$ & 175.1915833 & -26.4880000 & 2.1557 & $11.17^{+0.07}_{-0.09}$ & $125\pm44$ & $8.38\pm0.08$ & $<10.32$ & $<0.12$ & $<0.2$ & Yes \\
964 & 175.1869167 & -26.4779444 & 2.1435 & $9.87^{+0.17}_{-0.27}$ & $24\pm11$ & - & $<10.55$ & $<0.85$ & $<1.7$ & - \\
996$^a$ & 175.2447500 & -26.5062500 & 2.1657 & $10.72^{+0.11}_{-0.14}$ & $56\pm12$ & $8.74\pm0.08$ & $<10.98$ & $<0.65$ & $<1.8$ & - \\
999 & 175.2465833 & -26.4656389 & 2.1473 & $9.88^{+0.08}_{-0.10}$ & $59\pm8$ & $8.61\pm0.06$ & $<11.24$ & $<0.96$ & $<3.0$ & - \\
1046 & 175.2409583 & -26.4932778 & 2.1642 & $10.46^{0.12}_{0.17}$ & $136\pm31$ & - & $11.35\pm0.02$ & $0.89^{+0.04}_{-0.04}$ & $1.7\pm0.5$ & - \\ 
\hline 
\noalign{\vskip 0.0cm}
\noalign{\vskip 0.0cm}
\end{tabular}
\label{T:BigTable}
\end{sidewaystable*}

\renewcommand{\arraystretch}{1.5}
\begin{sidewaystable*}
\begin{tabular}{cccccccccccccccccc}
\hline
\noalign{\vskip 0.0cm}
ID & RA  & DEC   & z  &  $\mathrm{\log\,M_*/M_\odot}$  & SFR & $\mathrm{12+\log(O/H)}$ & $\mathrm{\log\,M_{mol}/M_\odot}$ & $\mathrm{f_{gas}}$ & $\mathrm{\tau_{dep}}$ & X-ray \\
   & (J2000) & (J2000) & & & ($\mathrm{M_\odot/yr}$) & & & & ($\mathrm{Gyr}$) & \\
\noalign{\vskip 0.1cm}
\hline
1047 & 175.2412917 & -26.4934167 & 2.1703 & $10.67^{+0.09}_{-0.12}$ & $84\pm6$ & - & $<11.06$ & $<0.71$ & $<1.4$ & Yes \\
1054 & 175.2408750 & -26.5133611 & 2.1644 & $11.16^{+0.08}_{-0.09}$ & $189\pm53$ & $8.66\pm0.14$ & $11.24\pm0.01$ & $0.55^{+0.07}_{-0.06}$ & $0.9\pm0.3$ & - \\
1066$^a$ & 175.2390833 & -26.4937500 & 2.1663 & $10.30^{+0.11}_{-0.14}$ & $23\pm6$ & $8.71\pm0.10$ & $<11.12$ & $<0.87$ & $<5.7$ & - \\ 
1071 & 175.2379583 & -26.4744167 & 2.1576 & $10.19^{+0.14}_{-0.20}$ & $26\pm20$ & - & $<10.61$ & $<0.72$ & $<1.5$ & - \\ 
1139 & 175.2300833 & -26.5119167 & 2.1447 & $9.68^{+0.13}_{-0.18}$ & $37\pm9$ & $8.39\pm0.20$ & $<11.42$ & $<0.98$ & $<7.1$ & - \\ 
1154 & 175.2299167 & -26.4783333 & 2.1630 & $10.39^{+0.08}_{-0.10}$ & $44\pm8$ & $8.56\pm0.12$ & $<10.87$ & $<0.75$ & $<1.7$ & - \\ 
1162$^a$ & 175.2272917 & -26.4732500 & 2.1610 & $10.73^{+0.09}_{-0.11}$ & $127\pm2$ & $8.71\pm0.08$ & $10.99\pm0.03$ & $0.64^{+0.07}_{-0.06}$ & $0.8\pm0.3$ & - \\ 
1181 & 175.2281250 & -26.4676111 & 2.1520 & $10.83^{+0.08}_{-0.10}$ & $166\pm55$ & - & $<11.05$ & $<0.62$ & $<0.7$ & - \\ 
1216 & 175.223625 & -26.5016111 & 2.1617 & $10.02^{+0.07}_{-0.09}$ & $61\pm5$ & - & $11.27\pm0.03$ & $0.94^{+0.02}_{-0.02}$ & $3.0\pm0.9$ & - \\ 
1284 & 175.2191667 & -26.5002500 & 2.1569 & $10.21^{+0.13}_{-0.19}$ & $40\pm14$ & - & $<10.54$ & $<0.68$ & $<0.9$ & - \\ 
1300 & 175.2135833 & -26.4940833 & 2.1531 & $10.90^{+0.09}_{-0.11}$ & $149\pm31$ & $8.62\pm0.06$ & $11.14\pm0.02$ & $0.64^{+0.08}_{-0.08}$ & $0.9\pm0.3$ & - \\ 
1316 & 175.2148333 & -26.4960833 & 2.1575 & $10.08^{+0.02}_{-0.02}$ & $11\pm1$ & $8.55\pm0.10$ & $<10.48$ & $<0.72$ & $<2.7$ & - \\ 
1385 & 175.2090833 & -26.4891389 & 2.1553 & $9.65^{+0.09}_{-0.11}$ & $24\pm6$ & - & $<11.07$ & $<0.96$ & $<5.0$ & - \\ 
1420 & 175.2057500 & -26.4858889 & 2.1661 & $9.34^{+0.15}_{-0.23}$ & $14\pm3$ & - & $<10.17$ & $<0.87$ & $<1.1$ & - \\ 
1440 & 175.2003750 & -26.4864722 & 2.1415 & $9.46^{+0.17}_{-0.28}$ & $16\pm4$ & - & $<10.54$ & $<0.92$ & $<1.7$ & - \\
1498 & 175.1998750 & -26.4804722 & 2.1628 & $10.12^{+0.12}_{-0.17}$ & $51\pm13$ & - & $<10.51$ & $<0.71$ & $<0.6$ & - \\ 
1501 & 175.1997500 & -26.4850833 & 2.1568 & $11.00^{+0.10}_{-0.13}$ & $245\pm94$ & - & $<10.68$ & $<0.33$ & $<0.2$ & Yes \\ 
28$^c$ & 175.2112500 & -26.4926667 & 2.1532 & $10.93^{+0.08}_{-0.10}$ & $141\pm39$ & - & $11.25\pm0.02$ & $0.68^{+0.06}_{-0.07}$ & $1.3\pm0.4$ & - \\ 
\hline 
\noalign{\vskip 0.0cm}
\noalign{\vskip 0.0cm}
\end{tabular}
\tablefoot{\\
\tablefoottext{a}{These objects are labeled as AGN candidates due to displaying $\log$([N{\sc{ii}}]/H$\alpha$)$\geq-0.35$ in \cite{PerezMartinez23}.}\\
\tablefoottext{b}{These objects are identified as type-1 AGN in \citealt{PerezMartinez23} due to their broad $\mathrm{H\alpha}$ component ($\sigma>700$ km\,s$^{-1}$). SFRs were computed from the $\mathrm{H\alpha}$ narrow-band component.}\\
\tablefoottext{c}{The ID of this HAE follow the scheme of \cite{Shimakawa18b} where it was first identified.}
}
\end{sidewaystable*}

\begin{sidewaystable*}
\caption{List of protocluster (PCL), protocluster core (PCL-core), cluster (CL), and field samples with CO(1-0) emitters from the literature at $\mathrm{z>1.5}$. The table columns display the name of the observed field, the redshift of the sample, the number ($\mathrm{N}$) of detected CO(1-0) sources, their median CO(1-0) luminosity ($\widetilde{L'_{\mathrm{CO(1-0)}}}$) and standard deviation, the area surveyed, the facility used for the observations, and the main reference for each sample.}
\begin{tabular}{cccccccc}
\hline
\noalign{\vskip 0.0cm}
Name & Type & Redshift & N & $\widetilde{L'_{\mathrm{CO(1-0)}}}$ & Area & Facility & Reference \\
     &      &          &   & ($\mathrm{10^{10}\,K\,km\,s^{-1}\,pc^2}$) & ($\mathrm{arcmin^2}$) &  &  \\
\noalign{\vskip 0.0cm}
\hline 
\hline 
PKS\,1138-262 & PCL & 2.16  & 46  & $3.5\pm2.2$ & 25.8 & ATCA  & \cite{Jin21}\\
CLJ\,1001+0220$^a$ & PCL & 2.51  & 14  & $2.2\pm4.2$ & 5.1\tablefootmark{a} & VLA  & \cite{Wang18}\\
CLJ\,1001+0220 & PCL &  2.51  & 8  & $1.7\pm1.4$ & 2.6 & VLA  & \cite{Champagne21}\\
USS\,1558-003 & PCL & 2.53  & 3  & $1.5\pm1.0$  & 1.5 & VLA  & \cite{Tadaki14}\\
HXMM20 & PCL-core & 2.60  & 5  & $5.0\pm2.6$ & 1.6 & VLA & \cite{Gomez-Guijarro19}\\
XMM-LSS\,J02182-05102 & CL & 1.62  & 2 & $1.7\pm1.3$ & 0.8 & VLA & \citet{Rudnick17}\\
\noalign{\vskip 0.0cm}
\hline
COSMOS & Field & $\mathrm{1.6<z<3.0}$  & 10 & $4.8\pm2.2$ & - & VLA  & \citet{Kaasinen19}\\
HUDF & Field & $\mathrm{2.0<z<2.7}$ & 6  & $1.7\pm1.0$ & 2.4 & VLA & \citet{Riechers20b}\\
COSMOS, GOODS-N  & Field & $\mathrm{2.0<z<2.8}$ & 51 & $2.9\pm3.4$ & - & VLA & \citet{Pavesi18} \\
HDF & Field & $\mathrm{2.2<z<2.5}$ & 5 & $7.6\pm3.5$ & 4.3 & VLA & \citet{Ivison11}\\
COSMOS, UDS, GOODS-N, EGS & Field & $\mathrm{2.2<z<4.4}$ & 13 & $10.6\pm3.7$  & - & VLA & \citet{FriasCastillo23}\\
\noalign{\vskip 0.0cm}
\hline
\end{tabular}
\tablefoot{\\
\tablefoottext{a}{\cite{Wang18} detect sources within twice the full width at half power (FWHP) per pointing and thus, its surveyed area takes this into account. Other works cited in this table restrict their surveyed areas to a single FWHP per pointing.}\\
}
\label{T:Littable}
\end{sidewaystable*}

\end{appendix}

\end{document}